\documentclass[aps,onecolumn,nofootinbib]{revtex4}

\usepackage{epsfig}
\usepackage{amsmath}
\usepackage{amsfonts}
\usepackage{amssymb}
\usepackage{graphicx}
\usepackage{colordvi}

\begin{document}

\def\pp{{\, \mid \hskip -1.5mm =}}
\def\cL{{\cal L}}
\def\be{\begin{equation}}
\def\ee{\end{equation}}
\def\bea{\begin{eqnarray}}
\def\eea{\end{eqnarray}}
\def\beq{\begin{eqnarray}}
\def\eeq{\end{eqnarray}}
\def\tr{{\rm tr}\, }
\def\nn{\nonumber \\}
\def\e{{\rm e}}
\def\de{\partial}
\def\lm{{\ell m}}


\title{ Black holes and stellar structures in $f(R)$-gravity}
\author{Mariafelicia De Laurentis\thanks{E-mail address: felicia@na.infn.it}  and Salvatore Capozziello\thanks{E-mail address: capozziello@na.infn.it}\\
{\it Dipartimento di Scienze Fisiche, Universit\'a
di Napoli {}``Federico II'', \\ and
\\ INFN Sezione  di Napoli, \\ Compl. Univ. di
Monte S. Angelo, Edificio G, Via Cinthia, I-80126, Napoli, Italy.}}



%


\begin{abstract}
In the last ten years, increasing attention has been devoted  to Extended Theories of Gravity with the aim to understand several  cosmological and astrophysical issues  such as the today observed accelerated expansion of the universe and the presence of Dark Matter in self-gravitating structures. Some of these models  assume modifications of General Relativity by adding higher order terms of  curvature invariants like the Ricci scalar $R$,  the Ricci tensor $R_{\mu\nu}$ and the Riemann  tensors $R_{\mu\nu\lambda\sigma}$, or the presence of suitable scalar fields like the former Brans-Dicke theory. It is therefore  natural to ask for  black hole solutions in this context since, on the one hand,  black holes signatures may be the test-bed to compare new models  to the Einstein gravity; on the other hand, they may lead to rule out  models which disagree with observations. 
Although black holes are one of the most striking predictions of General Relativity, they remain one of its least tested concepts. Electromagnetic observations  allow indirectly   to infer their existence, but direct evidences  remains elusive. In the next decade, data coming from very long-baseline interferometry  and gravitational wave detectors should allow  to image and study black holes in detail. Such observations will test General Relativity in the  non-linear and strong-field regimes where data are currently lacking.
Testing strong-field  features of General Relativity is of utmost importance to physics and astrophysics as a whole. This is because the  black holes solutions, such as the Schwarzschild and Kerr metrics, enter several calculations, including accretion disk structure, gravitational lensing, cosmology and gravitational waves theory. These  black hole solutions could indicate  strong-field departures from General Relativity with deep implications for the still unknown fundamental theory of gravity.
Beside the physical interest, black hole solutions represent a very active area for mathematical physics investigations.  
Here we review   the problem of black holes  in  a particular class of Extended Theories of Gravity, the so called $f(R)$-gravity, discussing some resolution techniques, obtaining exact solutions and comparing results with standard General Relativity. Furthermore, we discuss the problems of hydrostatic equilibrium and stellar structure in the context of $f(R)$-gravity showing that new features could emerge. The observation of such features could both explain the physics of exotic self-gravitating objects and 
constitute a signature for Extended Theories of Gravity.
\end{abstract}

\date{\today}


\maketitle

\section{Introduction}

The issue to extend General Relativity (GR)
 has recently become dramatically urgent due to  the missing matter problem at all astrophysical scales and the accelerating behavior of cosmic fluid, detected by Super Novae Ia used as standard candles. Up to now, no final answer on new particles has been given at fundamental level so Dark Energy and Dark Matter  constitute a puzzle to be solved in order to achieve a self-consistent picture of the observed Universe. $f(R)$-gravity, where $f(R)$ is a generic function of the Ricci scalar $R$,  comes into the game as a straightforward extension of GR where further geometrical degrees of freedom are considered instead of searching for new material ingredients \cite{f(R)-cosmo}. From an epistemological  point of view, the action of gravity  is not selected {\it a priori}, but it could be "reconstructed", in principle, by matching consistently the observations \cite{Capozz2, GRGrev,FoP}. This approach can be adopted considering any function of the curvature invariants as $R_{\mu\nu}R^{\mu\nu}$, $R\Box R$ and so on.
Several of these extended models  reproduce Solar System
tests so they are not in conflict with GR  experimental results  but actually extend them enclosing new features that could be, in principle, observed\cite{Capozziello8,Capozziello9, pogosian}.

From a genuine mathematical point of view, Extended Theories of Gravity pose the problem to recover the well-established results of GR as the initial value problem \cite{cauchy},  the stability of solutions  and, in particular, the issue of finding out new solutions.  As it is well known, beside cosmological solutions, spherically and axially symmetric solutions play a fundamental role in several astrophysical problems ranging from black holes to active galactic nuclei. Extended gravities, to be consistent with  results of GR, should comprise solutions like Schwarzschild and Kerr ones but could present, in general,  new solutions of  physical interest. Due to this reason, methods to find out exact and approximate solutions are particularly relevant in order to check if observations can be framed in Extended Theories of Gravity \cite{noether}.

Recently, the interest in spherically symmetric solutions of $f(R)$-gravity is growing up. In  \cite{Multamaki},  solutions in vacuum
have been found considering relations among functions that define the spherical metric or
imposing a costant Ricci curvatue scalar. The authors have reconstructed  the form of some $f(R)$-models,
discussing their physical relevance. In \cite{Multamaki1}, the
same authors have discussed static spherically symmetric  solutions,  in presence of perfect
fluid matter, adopting the metric formalism. They have
shown that a given matter distribution is not capable of globally determining
the functional form of $f(R)$. Others authors have discussed  in
details the spherical symmetry  of $f(R)$-gravity considering also
the relations with the weak field limit.  Exact solutions are
obtained for constant Ricci curvature scalar and for Ricci scalar
depending on the radial coordinate. In particular, it can be considered
how to obtain results  consistent with GR assuming the well-known post-Newtonian and post-Minkowskian limits as consistency checks
\cite{arturo}.

As we will discuss below,   a general method to find out
rotating black hole solutions  can be achieved by performing a complex coordinate transformation
on the spherical black hole metrics. Since the discovery of the Kerr solution \cite{pap:kerr},  many attempts have been made to find a
physically reasonable interior matter distribution that may be
considered as its source. For a review on these approaches
see \cite{pap:dmm, pap:ak}. Though much progress has been made, results have
been generally disappointing. As far as we know,  nobody  has
obtained a physically satisfactory interior solution. This seems
surprising given the success of matching internal spherically symmetric solutions to the Schwarzschild metric. The problem is
 that the loss of a degree of symmetry makes the
derivation of analytic results  much more difficult.  Severe
restrictions are placed on  the interior metric by maintaining that it
must be joined smoothly to the external axially symmetric metric.
Further restrictions are placed on the  interior solutions to ensure that they correspond to
physical objects.

Furthermore
since the axially symmetric metric has no radiation field associated with it,
its source should  be also non-radiating. This places even
further constraints on the structure of the interior
solution  \cite{bk:dk}. Given the strenuous nature of these limiting
conditions, it  is not surprising to learn that  no
satisfactory solution to the problem of finding  sources
for the Kerr metric has been obtained. In general,  the failure is  due to  internal
structures whose physical properties are unknown. This shortcoming makes hard to find consistent
 boundary conditions.

Newman and Janis
showed that it  is possible to  obtain   rotating  solutions (like the
Kerr metric) by making an elementary complex transformation on the
Schwarzschild solution~\cite{pap:nj1}. This same method has been
used to obtain a new stationary and axially symmetric
solution known as the
Kerr-Newman metric~\cite{pap:nj2}. The Kerr-Newman space-time is
associated to the exterior geometry of a rotating massive and
charged black-hole. For a  review on the Newman-Janis method
 to obtain both the Kerr and Kerr-Newman metrics
see~\cite{bk:ier,turolla}.

By means of very elegant
mathematical arguments, Schiffer et al.~\cite{pap:mms} have  given a rigorous
proof to show how the Kerr metric can be derived starting from a complex
transformation on the Schwarzschild solution. We will not go
into  details of this demonstration, but point out  that
 the proof  relies on two main assumptions. The first  is that  metric
belongs to the same algebraic class of the Kerr-Newman solution,
namely the Kerr-Schild class \cite{pap:gcd}. The second assumption is
that  metric corresponds to an empty solution of the Einstein
field equations   \cite{pap:mms}. It is clear, by the
generation of the Kerr-Newman metric, that all the components of
the stress-energy tensor need to be non-zero for the Newman-Janis method to be
successful.   Such a transformation
can be  extended to $f(R)$-gravity  as discussed in \cite{axially}.

On the other hand, the strong gravity regime   is
another way to check the viability of these theories \cite{psaltis}. In general the formation and the evolution of stars can be considered suitable test-beds for
Extended Theories of Gravity.  Considering the case of $f(R)$-gravity,  divergences
stemming from the functional form of $f(R)$ may prevent the existence of
relativistic stars in these theories \cite{briscese}, but thanks
to the chameleon mechanism, introduced by   Khoury and  Weltman \cite{weltman}, the possible problems jeopardizing the existence
of these objects may be avoided \cite{tsu}. Furthermore, there are
also numerical solutions corresponding to static star configurations with
 strong gravitational fields \cite{babi} where the choice of the
equation of state  is crucial for the existence of solutions.

It is also important to stress that $f(R)$-gravity  has interesting applications also in stellar astrophysics and could contribute to solve
several puzzles related to observed peculiar objects (e.g.  stars in the instability strips, protostars, etc. \cite{Cooney,Hu}),
structure and star formation \cite{poly, Chang}.
Furthermore some observed stellar systems  are incompatible with the
standard models of stellar structure. We refer to anomalous neutron stars, the so called 
"magnetars"  \cite{mag}  with masses larger than their expected Volkoff mass.
It  seems that, on particular length scales, the gravitational force is larger
or smaller than the corresponding GR value.
For example, a modification of the Hilbert-Einstein Lagrangian, consisting of  $R^2$ terms,
enables a major attraction while a $R_{\alpha\beta}R^{\alpha\beta}$ term gives a repulsive  
contribution \cite{stabile_2}. Understanding 
on which scales the modifications to GR are working or what is the weight of corrections 
to gravitational potential is a crucial point that could confirm or rule out these extended approaches to gravitational interaction. 

This Chapter is organized as follow. 
 In the Sec.\ref{due}, we
introduce  the $f(R)$-gravity action,  the field
equations and  give some general remarks
on spherical symmetry. In Sec ~\ref{tre},   a summary is given on  the Noether Simmetry Approach  \cite{noether}. This technique is extremely useful to
 find  exact solutions. In particular, we find spherically symmetric black hole solutions  for $f(R)$-gravity.
In Sec ~\ref{quattro},
 we review the Newman-Janis method  to obtain rotating solutions
starting from spherically symmetric ones. The resulting metric is written in terms of
two arbitrary functions.  A further suitable   coordinate
transformation allows to write the  metric  in the so called
Boyer-Lindquist  coordinates. Such a transformation makes the physical
interpretation much clearer and reduces the amount of algebra
required to calculate the  metric properties.
In Sec.\ref{cinque}, the Newman-Janis method  is applied to a spherically symmetric exact solution, previously derived
 by the Noether Symmetry, and an axially symmetric exact
solution is obtained. This result shows  that the Newman-Janis method works also in $f(R)$-gravity.
Physical applications of the result are  also discussed. 

In Sec.\ref{sei}, we review  the classical hydrostatic problem for stellar structures. In Sec. \ref{sette} we derive the  
Newtonian limit of $f(R)$-gravity obtaining the modified Poisson equation. The modified Lan\'{e}-Emden equation is obtained in Sec. \ref{otto}  and its structure is compared  with respect to the standard one. In Sec.\ref{nove}, we show how analytical solutions of standard Lan\'{e}-Emden equation can be compared  with those  perturbatively obtained  from $f(R)$-gravity.  
In order to apply the above results,  in Section \ref{dieci} the classical theory  of gravitational collapse
for dust-dominated systems is summarized.  In Section \ref{undici}, we  discuss the weak field   limit of $f(R)$-gravity obtaining corrections to the standard
Newtonian potential that can be figured out as two Newtonian potentials contributing to the dynamics. In Section \ref{dodici} we
recover the dispersion relation and Jeans mass limit\cite{jeans}. Some self-gravitating dust  system are discussed in this approach.
The difference between GR and $f(R)$-gravity are put in evidence, in particular the Jeans mass profiles with respect to the temperature.
We report a catalogue of observed molecular clouds in order to compare the classical Jeans mass  to the $f(R)$-one. Finally, in Section \ref{discuss}, we discuss the results and draw conclusions.


\section{Spherical symmetry in  $f(R)$-gravity }
\label{due}
Let us start by discussing exact solutions in  $f(R)$-gravity with spherical symmetry. 
As we will see, a  crucial role  is played by the relation between the metric potentials
 and the Ricci scalar that can be regarded as a constraint  assuming the form of a Bernoulli equation.
 
Let us consider an analytic function $f(R)$ of the Ricci scalar
$R$ in four dimensions. The variational principle for this action is:

\begin{equation}\label{fRaction}
\delta\int
d^4x\sqrt{-g}\biggl[f(R)+\mathcal{X}\mathcal{L}_m\biggr]\,=\,0
\end{equation}
where ${\displaystyle \mathcal{X}=\frac{8\pi G}{c^4}}$,
$\mathcal{L}_m$ is the standard matter Lagrangian and $g$ is the
determinant of the metric\footnote{We are adopting the convention $c=1$.  The convention
$R_{\mu\nu}={R^\rho}_{\mu\rho\nu}$ for the Ricci tensor and
${R^\alpha}_{\beta\mu\nu}=\Gamma^\alpha_{\beta\nu,\mu}-...$, for
the Riemann tensor. Connections are Levi-Civita \,:
\begin{equation}
\Gamma^\mu_{\alpha\beta}=\frac{1}{2}g^{\mu\rho}(g_{\alpha\rho,\beta}+g_{\beta\rho,\alpha}-g_{\alpha\beta,\rho})\,.\nonumber\\
\end{equation} The signature is $(+\,-\,-\,-)$.}.

By varying  with respect to the metric, we obtain the field
equations \footnote{It is possible to take into account also the
Palatini approach in which  the metric $g$ and the connection
$\Gamma$  are considered independent variables (see for example
\cite{palatini}). Here we will consider the Levi-Civita connection
and will use the metric approach. See \cite{GRGrev, palatini,ACCF} for a
detailed comparison between the two pictures.}

\begin{equation}\label{HOEQ}
\left\{\begin{array}{ll}H_{\mu\nu}=f'(R)R_{\mu\nu}-\frac{1}{2}f(R)g_{\mu\nu}-f'(R)_{;\mu\nu}+g_{\mu\nu}\Box
f'(R)\,=\,\mathcal{X}T_{\mu\nu}\\\\H\,=\,g^{\rho\sigma}H_{\rho\sigma}=3\Box
f'(R)+f'(R)R-2f(R)\,=\,\mathcal{X}T\end{array}\right.
\end{equation}
where
$T_{\mu\nu}\,=\,\displaystyle\frac{-2}{\sqrt{-g}}\frac{\delta(\sqrt{-g}\mathcal{L}_m)}{\delta
g^{\mu\nu}}$ is the energy-momentum tensor of standard fluid matter and the second equation is the trace. The
most general spherically symmetric solution can be written as
follows\,:

\begin{equation}\label{me0}
ds^2\,=\,m_1(t',r')dt'^2+m_2(t',r')dr'^2+m_3(t',r')dt'dr'+m_4(t',r')d\Omega\,,
\end{equation}
where $m_i$ are functions of the radius $r'$ and of the time $t'$.
 $d\Omega$ is the solid angle. We can consider a
coordinate transformation that maps the metric (\ref{me0}) in a
new one where the off\,-\,diagonal term vanishes and
$m_4(t',r')\,=\,-r^2$, that is\footnote{This condition allows to
obtain the standard definition of the circumference with  radius
$r$.}\,:

\begin{equation}\label{me}
ds^2\,=\,g_{tt}(t,r)dt^2-g_{rr}(t,r)dr^2-r^2d\Omega\,.
\end{equation}
This  expression can be considered, without loss of generality, as
the most general definition of a spherically symmetric metric
compatible with a pseudo\,-\,Riemannian manifold  without torsion.
Actually, by inserting this metric into the field Eqs.
(\ref{HOEQ}), one obtains\,:

\begin{equation}\label{fe4}
\left\{\begin{array}{ll}f'(R)R_{\mu\nu}-\frac{1}{2}f(R)g_{\mu\nu}+\mathcal{H}_{\mu\nu}\,=\,\mathcal{X}T_{\mu\nu}\\\\
f'(R)R-2f(R)+\mathcal{H}\,=\,\mathcal{X}T\end{array}\right.
\end{equation}
where the two quantities $\mathcal{H}_{\mu\nu}$ and $\mathcal{H}$
read\,:
\begin{eqnarray}\label{highterms1}
\mathcal{H}_{\mu\nu}\,&=&\,-f''(R)\biggl\{R_{,\mu\nu}-\Gamma^t_{\mu\nu}R_{,t}-\Gamma^r_{\mu\nu}R_{,r}-
g_{\mu\nu}\biggl[\biggl({g^{tt}}_{,t}+\nonumber\\&&+g^{tt}
\left(\ln\sqrt{-g}\right)_{,t}\biggr)R_{,t}+\biggl({g^{rr}}_{,r}+g^{rr}\left(\ln\sqrt{-g}\right)_{,r}\biggr)R_{,r}+\nonumber\\&&+g^{tt}R_{,tt}
+g^{rr}R_{,rr}\biggr]\biggr\}-f'''(R)\biggl[R_{,\mu}R_{,\nu}-g_{\mu\nu}\biggl(g^{tt}{R_{,t}}^2+g^{rr}
{R_{,r}}^2\biggr)\biggr]
\end{eqnarray}
\begin{eqnarray}
\label{highterms2}
\mathcal{H}\,&=&\,g^{\sigma\tau}\mathcal{H}_{\sigma\tau}\,=\,3f''(R)\biggl[\biggl({g^{tt}}_{,t}+g^{tt}
\left(\ln\sqrt{-g}\right)_{,t}\biggr)R_{,t}+\nonumber\\&&+\biggl({g^{rr}}_{,r}+g^{rr}\left(\ln\sqrt{-g}\right)_{,r}\biggr)R_{,r}+g^{tt}R_{,tt}
+g^{rr}R_{,rr}\biggr]+
3f'''(R)\biggl[g^{tt}{R_{,t}}^2+g^{rr}{R_{,r}}^2\biggr]\,.\nonumber\\
\end{eqnarray}
Our task is now to find out
exact spherically symmetric solutions.

In the case of time-independent metric, i.e., $g_{tt}\,=\,a(r)$
and $g_{rr}\,=\,b(r)$, the Ricci scalar  can
be recast as a Bernoulli equation of index two with respect to
the metric potential $b(r)$\, (see \cite{arturo} for details):

\begin{eqnarray}\label{eqric}
&& b'(r)+\biggl\{\frac{r^2a'(r)^2-4a(r)^2-2ra(r)[2a(r)'+ra(r)'']}{ra(r)[4a(r)
+ra'(r)]}\biggr\}b(r)+\nonumber\\&&+\biggl\{\frac{2a(r)}{r}\biggl[\frac{2+r^2R(r)}{4a(r)+ra'(r)}\biggr]\biggr\}b(r)^2\,=\,0\,.
\end{eqnarray}
where $R\,=\,R(r)$ is the Ricci scalar. A general solution of
(\ref{eqric}) is:

\begin{equation}\label{gensol}
b(r)\,=\,\frac{\exp[-\int dr\,h(r)]}{K+\int dr\,l(r)\,\exp[-\int
dr\,h(r)]}\,,
\end{equation}
where $K$ is an integration constant while $h(r)$ and $l(r)$ are
 two functions that, according to Eq.(\ref{eqric}), define the coefficients of
the quadratic and the linear term with respect to $b(r)$
\cite{bernoulli}. We can fix  $l(r)\,=\,0$; this choice allows to find out
solutions with a Ricci scalar scaling as ${\displaystyle
-\frac{2}{r^2}}$ in term of the radial coordinate. On the other
hand, it is not possible to have $h(r)\,=\,0$ since, otherwise, we
 get imaginary solutions. A particular consideration deserves
the limit $r\rightarrow\infty$. In order to achieve a
gravitational potential $b(r)$ with the correct Minkowski limit,
both $h(r)$ and $l(r)$ have to go to zero at infinity, provided that the
quantity $r^2R(r)$ turns out to be constant: this result implies
$b'(r)=0$, and, finally, also the metric potential $b(r)$ has  a
correct Minkowski limit.

In general,  if we ask for the asymptotic flatness of the metric
as a feature of the theory,  the Ricci scalar has to evolve to
infinity as $r^{-n}$ with $n\geqslant 2$. Formally, it has to be:

\begin{equation}\label{condricc}
\lim_{r\rightarrow\infty}r^2R(r)\,=\,r^{-n}\,,
\end{equation}
with $n\in\mathbb{N}$. Any other behavior of the Ricci scalar
could affect the requirement to achieve a correct asymptotic flatness.

The case of constant curvature is equivalent to GR with a
cosmological constant and the solution is time independent. This
result is well known (see, for example, \cite{barrottew}) but we
report, for the sake of completeness, some considerations related
with it in order to deal with more general cases where a radial
dependence for the Ricci scalar is supposed.  If the scalar curvature is constant
($R\,=\,R_0$),  field Eqs.(\ref{fe4}), being
$\mathcal{H}_{\mu\nu}\,=\,0$,  reduce to:

\begin{equation}\label{fe2}
\left\{\begin{array}{ll}f'_0R_{\mu\nu}-\frac{1}{2}f_0g_{\mu\nu}\,=\,\mathcal{X}T_{\mu\nu}\\\\
f'_0R_0-2f_0\,=\,\mathcal{X}T\end{array}\right.
\end{equation}
where $f(R_0)=f_0$, $f'(R_0)=f'_0$. A general solution, when one considers a stress-energy tensor
of perfect-fluid $T_{\mu\nu}\,=\,(\rho+p)u_\mu u_\nu-pg_{\mu\nu}$,
is

\begin{equation}
ds^2\,=\,\biggl(1+\frac{k_1}{r}+\frac{q\mathcal{X}\rho-\lambda}{3}r^2\biggr)dt^2-\frac{dr^2}{1+\frac{k_1}{r}+
\frac{q\mathcal{X}\rho-\lambda}{3}r^2}-r^2d\Omega\,.
\end{equation}
when $p\,=\,-\rho$, $\lambda=-\frac{f_0}{2f'_0}$ and
$q^{-1}=f'_0$. This result means that any $f(R)$-model, in the
case of constant curvature, exhibits solutions
with de Sitter-like behavior. This is
one of the reasons why the dark energy issue can be addressed using
these theories \cite{f(R)-cosmo}.

If $f(R)$ is analytic, it is possible to write the series:
\begin{equation}\label{f}
f(R)\,=\,\Lambda+\Psi_0R+\Psi(R)\,,
\end{equation}
where $\Psi_0$ is a coupling constant, $\Lambda$ plays the role of
the cosmological constant and $\Psi(R)$ is a generic analytic
function of $R$ satisfying the condition

\begin{equation}\label{psi}
\lim_{R\rightarrow 0}R^{-2}\Psi(R)\,=\,\Psi_1\,,
\end{equation}
where $\Psi_1$ is a constant. If we neglect the cosmological
constant $\Lambda$ and $\Psi_0$ is set to zero, we obtain a new
class of theories which, in the limit $R\rightarrow{0}$, does not
reproduce GR (from  Eq.(\ref{psi}), we have $\lim_{R\rightarrow 0}
f(R)\sim R^2$). In such a case, analyzing the whole set of
Eqs.(\ref{fe2}), one can observe that both zero and constant $\neq
0$ curvature solutions are possible. In particular, if
$R\,=\,R_0\,=\,0$ field equations are solved for any form of
gravitational potential entering the spherically symmetric
background,  provided that the Bernoulli Eq. (\ref{eqric}),
relating these functions, is fulfilled for the particular case
$R(r)=0$. The  solutions are thus defined by the relation

\begin{equation}\label{gensol0}
b(r)\,=\,\frac{\exp[-\int
dr\,h(r)]}{K+4\int\frac{dr\,a(r)\,\exp[-\int
dr\,h(r)]}{r[a(r)+ra'(r)]}}\,,
\end{equation}
being $g_{tt}(t,r)=b(r)$ from Eq.(\ref{me}). In \cite{arturo},
some examples of $f(R)$-models admitting
solutions with constant$\neq 0$ or null scalar curvature are discussed.

\section{The Noether Symmetry Approach }
\label{tre}

Besides spherically symmetric solutions with constant curvature scalar, also solutions with the Ricci scalar depending on radial coordinate $r$ are possible in $f(R)$-gravity \cite{arturo}. Furthermore, spherically symmetric solutions
 can be achieved  starting from a
point-like $f(R)$-Lagrangian  \cite{noether}. Such a Lagrangian can be obtained by
imposing the spherical symmetry  directly in the   action
(\ref{fRaction}). As a consequence, the infinite number of degrees
of freedom of the original field theory will be reduced to a
finite number. The technique is based on the choice of a suitable
Lagrange multiplier defined by assuming the Ricci scalar, argument
of the function $f(R)$ in spherical symmetry.

Starting from the above considerations, a static
spherically symmetric metric can be expressed as

\begin{equation}\label{me2}
{ds}^2=A(r){dt}^2-B(r){dr}^2-M(r)d\Omega\,,
\end{equation}
and then the point-like $f(R)$ Lagrangian\footnote{Obviously, the above choices are recovered for $A(r)=a(r)$, $B(r)=b(r)$, and $M(r)=r^2$. Here we deal with $A,B,M$ as a set of coordinates in a configuration space.}  is

\begin{eqnarray}\label{lag2}
\mathcal{L}&=&-\frac{A^{1/2}f'}{2MB^{1/2}}{M'}^2-\frac{f'}{A^{1/2}B^{1/2}}A'M'-\frac{Mf''}{A^{1/2}B^{1/2}}A'R'+\nonumber\\&&-\frac{2A^{1/2}f''}{B^{1/2}}R'M'-A^{1/2}B^{1/2}[(2+MR)f'-Mf]\,,
\end{eqnarray}
which is canonical since only the configuration variables and
their first order derivatives with respect to the radial coordinate $r$ are present. Details of calculations are in \cite{noether}.
 Eq.
(\ref{lag2}) can be recast in a more compact form introducing the
matrix representation\,:
\begin{equation}\label{la}
\mathcal{L}={\underline{q}'}^t\hat{T}\underline{q}'+V
\end{equation}
where $\underline{q}=(A,B,M,R)$ and $\underline{q}'=(A',B',M',R')$
are the generalized positions and velocities associated to
$\mathcal{L}$.  It is easy to check
the complete analogy between the field equation approach and
point-like Lagrangian approach \cite{noether}.

In order to find out solutions for the  Lagrangian (\ref{lag2}),
we can search for symmetries related to cyclic variables and then
reduce dynamics. This approach allows, in principle, to select
$f(R)$-gravity models compatible with spherical symmetry. As a
general remark, the Noether Theorem states that conserved
quantities are related to the existence of cyclic variables into
dynamics \cite{arnold,marmo,morandi}.

It is worth noticing that the Hessian determinant of  Eq. (\ref{lag2}),
${\displaystyle \left|\left|\frac{\partial^2\mathcal{L}}{\partial
q'_i\partial q'_j}\right|\right|}$, is zero. This result clearly
depends on the absence of the generalized velocity $B'$ into the
point\,-\,like Lagrangian. As matter of fact, using a point-like
Lagrangian approach implies that the metric variable $B$ does not
contributes to dynamics, but the  equation of motion for $B$ has
to be considered as a further constraint equation. Then  the
Lagrangian (\ref{lag2})  has three  degrees of freedom and not
four, as one should expect  {\it a priori}.

Now, since the  equation of motion describing the evolution of the
metric potential $B$ does not depend on its derivative, it can be
explicitly solved in term of $B$ as a function of the other
coordinates\,:
\begin{equation}\label{eqb}
B=\frac{2M^2f''A'R'+2Mf'A'M'+4AMf''M'R'+Af'M'^2}{2AM[(2+MR)f'-Mf]}\,.
\end{equation}
By inserting Eq.(\ref{eqb}) into the Lagrangian (\ref{lag2}),  we
obtain a non-vanishing Hessian matrix removing the singular
dynamics. The new Lagrangian reads\footnote{Lowering  the
dimension of configuration space through the substitution
(\ref{eqb}) does not affect the  dynamics since $B$ is a
non-evolving quantity. In fact, inserting Eq. (\ref{eqb})
 into the dynamical equations given by (\ref{lag2}),  they
coincide with those derived by (\ref{lag2}).}
\begin{equation}\label{}\mathcal{L}^*= {\bf L}^{1/2}\end{equation}
with
\begin{eqnarray}\label{lag3}\nonumber
{\bf L}=\underline{q'}^t\hat{{\bf
L}}\underline{q'}=\frac{[(2+MR)f'-fM]}{M}[2M^2f''A'R'
+2MM'(f'A'+2Af''R') +Af'M'^2]\,.
\end{eqnarray}

If one assumes the spherical
symmetry, the role of the {\it affine parameter} is  played by the
coordinate radius $r$.  In this case, the configuration space is
given by $\mathcal{Q}=\{A, M, R\}$ and the tangent space by
$\mathcal{TQ}=\{A, A', M, M', R, R'\}$. On the other hand,
according to the Noether Theorem, the existence of a symmetry for
dynamics described by  Lagrangian (\ref{lag2})  implies a
constant of motion. Let us apply the Lie derivative to the
(\ref{lag2}), we have\footnote{From now on,  $\underline{q}$
indicates the vector $\{A,M,R\}$.}\,:
\begin{equation}\label{}
L_{\mathbf{X}}{\bf L}\,=\,\underline{\alpha}\cdot\nabla_{q}{\bf L
}+\underline{\alpha}'\cdot\nabla_{q'}{\bf L}
=\underline{q}'^t\biggl[\underline{\alpha}\cdot\nabla_{q}\hat{{\bf
L }}+ 2\biggl(\nabla_{q}\alpha\biggr)^t\hat{{\bf L
}}\biggr]\underline{q}'\,,
\end{equation}
that vanishes if the functions ${\underline{\alpha}}$ satisfy the
following system
\begin{equation}\label{sys}
\underline{\alpha}\cdot\nabla_{q}\hat{{\bf L}}
+2(\nabla_{q}{\underline{\alpha}})^t\hat{{\bf L
}}\,=\,0\,\longrightarrow\ \ \ \ \alpha_{i}\frac{\partial
\hat{{\bf L}}_{km}}{\partial
q_{i}}+2\frac{\partial\alpha_{i}}{\partial q_{k}}\hat{{\bf L
}}_{im}=0\,.
\end{equation}
Solving the system (\ref{sys})  means to find out the functions
$\alpha_{i}$ which assign the Noether vector \cite{arnold,marmo}. However the system
(\ref{sys}) implicitly depends on the form of $f(R)$ and then, by
solving it, we get also $f(R)$-models compatible with spherical
symmetry. On the other hand, by choosing the $f(R)$-form, we can
explicitly solve (\ref{sys}). As an example, one finds that the
system (\ref{sys}) is satisfied if we choose
\begin{equation}\label{solsy}f(R)\,=\,f_0 R^s\,,\ \ \ \ \  \underline{\alpha}=(\alpha_1,\alpha_2,\alpha_3)=
\biggl((3-2s)kA,\ -kM,\ kR\biggr)\,,
\end{equation}
with $s$ a real number, $k$ an integration constant and $f_0$ a
dimensional coupling constant\footnote{ The dimensions are given
by $R^{1-s}$ in terms of the Ricci scalar. For the sake of
simplicity, we will put $f_0=1$ in the forthcoming discussion.}.
This means that, for any $f(R)=R^s$, exists, at least, a Noether
symmetry and  a related constant of motion $\Sigma_{0}$\,:
\begin{eqnarray}\label{cm}\nonumber
\Sigma_{0}\,=\,\underline{\alpha}\cdot\nabla_{q'}{\bf L
}=2
skMR^{2s-3}[2s+(s-1)MR][(s-2)RA'-(2s^2-3s+1)AR']\,.\end{eqnarray}
A physical interpretation of $\Sigma_{0}$  is possible if one
gives an interpretation of this quantity in GR, that means for $f(R)=R$ and $s=1$. In other words, the above procedure has to be applied to the
Lagrangian of GR. We obtain the solution
\begin{equation}\label{solsygr}\underline{\alpha}_{GR}=(-kA,\
kM)\,.
\end{equation}
The functions $A$ and $M$ give the Schwarzschild solution, and
then the constant of motion acquires the standard form
\begin{equation}\label{cmgr}\Sigma_{0}= \frac{2GM}{c^2}\,.\end{equation}
In other words, in the case of Einstein gravity, the Noether
symmetry gives, as a conserved quantity,   the Schwarzschild radius
or  the mass of the gravitating system.  This result can be assumed as a consistency check.

In the general case, $f(R)=R^s$, the Lagrangian (\ref{lag2})
becomes
\begin{eqnarray}\label{}\nonumber {\bf L}&=&\frac{sR^{2s-3}[2s+(s-1)MR]}{M}[2(s-1)M^2A'R'+2MRM'A'+\nonumber\\&& 4(s-1)AMM'R'+ARM'^2]\,,\end{eqnarray}
and the expression (\ref{eqb}) for $B$ is
\begin{equation}\label{}B=\frac{s[2(s-1)M^2A'R'+2MRM'A'+4(s-1)AMM'R'+ARM'^2]}{2AMR[2s+(s-1)MR]}\end{equation}
As it can be easily checked, GR is recovered for $s=1$.

Using the constant of motion (\ref{cm}),  we solve in term of $A$
and obtain
\begin{equation}\label{}A=R^{\frac{2s^2-3s+1}{s-2}}\biggl\{k_1+\Sigma_{0}\int\frac{R^{\frac{4s^2-9s+5}{2-s}}dr}{2ks(s-2)M[2s+(s-1)MR]}\biggr\}\end{equation}
for $s\neq2$ and  $k_1$ an integration constant. For $s\,=\,2$,
one finds
\begin{equation}\label{}A=-\frac{\Sigma_{0}}{12kr^2(4+r^2R)RR'}\,.\end{equation}
These relations allow to find out general black hole solutions for the field
equations giving the dependence of the Ricci scalar on the radial
coordinate $r$. For example, a solution is found for
\begin{equation}\label{}
s=5/4\,,\ \ \ \ M=r^2\,,\ \ \ \ R= 5 r^{-2}\,,
\end{equation}
obtaining  the spherically symmetric space-time
\begin{equation}\label{sol_noe_2}ds^2=(\alpha+\beta
r)dt^2-\frac{1}{2}\frac{\beta r}{\alpha+\beta r}
dr^2-r^2d\Omega\,,\end{equation}
where $\alpha$ is a combination of  $\Sigma_0$ and $k$ and $\beta=k_1$.
In principle, the same procedure can be worked out any time Noether symmetries are identified.
Our task is now to show how, from  a spherically symmetric solution, one can  generate an axially symmetric solution adopting the Newman-Janis procedure that works in GR.  In general, the approach is not immediately straightforward since, as soon as $f(R)\neq R$, we are dealing with fourth-order field equations which have, in principle, different existence theorems and boundary conditions. However, the existence of the Noether symmetry guarantees the consistency of the chosen $f(R)$-model with the field equations.

\section{Axial symmetry derived from spherical symmetry}
\label{quattro}

We want to show now how it is possible to obtain an rotating solution starting from a spherically symmetric one adopting the method developed by Newman and Janis in GR. Such an algorithm can be applied to a
static spherically symmetric  metric considered as a``seed'' metric.
Let us recast   the  spherically symmetric metric (\ref{me}) in the form

\begin{equation}
ds^2 = e^{2\phi(r)}dt^2 - e^{2\lambda(r)}dr^2 - r^2d\Omega,
\label{eqn:ssm}
\end{equation}
with $g_{tt}(t,r)\,=\,e^{2\phi(r)}$ and
$g_{rr}(t,r)\,=\,e^{2\lambda(r)}$. Such a form is suitable for the considerations below. Following Newman and Janis,
Eq.~(\ref{eqn:ssm}) can be written in the so called
Eddington--Finkelstein coordinates $(u,r,\theta,\phi)$, i.e. the
$g_{rr}$ component is eliminated by a change of coordinates and a
cross term is introduced \cite{gravitation}. Specifically this is achieved by defining
the time coordinate as $dt = du + F(r)dr$ and setting $F(r)= \pm
e^{\lambda(r)-\phi(r)}$. Once such a transformation is performed,  the metric
(\ref{eqn:ssm})  becomes

\begin{equation}\label{nullelemn}
ds^2 = e^{2\phi(r)}du^2 \pm 2e^{\lambda(r)+\phi(r)}dudr -
r^2d\Omega.
\end{equation}
The surface $u\,=\,$ costant is a light cone starting
from the origin $r\,=\,0$. The metric tensor for the line element
(\ref{nullelemn}) in null-coordinates is

\begin{equation}\label{metrictensorcontro}
g^{\mu\nu} = \left(
\begin{array}{cccc}
0 & \pm e^{-\lambda(r)-\phi(r)} & 0 & 0     \\
\pm e^{-\lambda(r)-\phi(r)} & -e^{-2\lambda(r)} & 0 & 0 \\
0 & 0 & -1/r^2 & 0 \\
0 & 0 & 0 & -1/(r^2\sin^2{\theta})
\end{array}    \right).
\end{equation}
The matrix  (\ref{metrictensorcontro}) can be written in terms of a null
tetrad as

\begin{equation}\label{eq:gmet}
g^{\mu\nu} = l^\mu n^\nu + l^\nu n^\mu-m^\mu\bar{m}^\nu -
m^\nu\bar{m}^\mu,
\end{equation}
where $l^\mu$, $n^\mu$, $m^\mu$ and $\bar{m}^\mu$ are the vectors
satisfying the conditions

\begin{equation}l_\mu l^\mu\,=\,m_\mu m^\mu\,=\,n_\mu n^\mu\,=\,0,\,\,\,\,\,
l_\mu n^\mu\,=\,-m_\mu\bar{m}^\mu\,=\,1, \,\,\,\,\,l_\mu
m^\mu\,=n_\mu m^\mu\,=\,0\,.\end{equation} The bar indicates the
complex conjugation. At any point in space, the tetrad can be chosen
in the following manner: $l^\mu$ is the outward null vector
tangent to the cone, $n^\mu$ is the inward null vector pointing
toward the origin, and $m^\mu$ and $\bar{m}^\mu$ are the vectors
tangent to the two-dimensional sphere defined by constant $r$ and
$u$. For the spacetime (\ref{metrictensorcontro}), the tetrad null
vectors can be

\begin{equation}
\left\{\begin{array}{ll}l^\mu\,=\,\delta^\mu_1\\\\n^\mu\,=\,-\frac12
e^{-2\lambda(r)}\delta^\mu_1+
e^{-\lambda(r)-\phi(r)}\delta^\mu_0\\\\
m^\mu\,=\,\frac{1}{\sqrt{2}r}(\delta^\mu_2
+\frac{i}{\sin{\theta}}\delta^\mu_3)\\\\
\bar{m}^\mu\,=\,\frac{1}{\sqrt{2}r}(\delta^\mu_2
-\frac{i}{\sin{\theta}}\delta^\mu_3)\end{array}\right.
\end{equation}
Now we need to extend the set of coordinates
$x^\mu\,=\,(u,r,\theta,\phi)$ replacing the real
radial coordinate by a complex variable. Then the tetrad
null vectors become \footnote{It is worth noticing that a certain
arbitrariness  is present in the
complexification process of the functions $\lambda$ and $\phi$.  Obviously, we have to obtain the metric (\ref{metrictensorcontro})  as soon as $r\,=\,\bar{r}$.}

\begin{equation}\label{tetradvectors}
\left\{\begin{array}{ll}l^\mu\,=\,\delta^\mu_1\\\\n^\mu\,=\,-\frac12
e^{-2\lambda(r,\bar{r})}\delta^\mu_1+
e^{-\lambda(r,\bar{r})-\phi(r,\bar{r})}\delta^\mu_0\\\\
m^\mu\,=\,\frac{1}{\sqrt{2}\bar{r}}(\delta^\mu_2
+\frac{i}{\sin{\theta}}\delta^\mu_3)\\\\
\bar{m}^\mu\,=\,\frac{1}{\sqrt{2}r}(\delta^\mu_2
-\frac{i}{\sin{\theta}}\delta^\mu_3)\end{array}\right.
\end{equation}
A new metric is obtained by making a complex coordinates
transformation

\begin{equation}\label{transfo}x^\mu\rightarrow\tilde{x}^\mu=x^\mu+iy^\mu(x^\sigma)\,,\end{equation}
where $y^\mu(x^\sigma)$ are analityc functions of the real
coordinates $x^\sigma$, and simultaneously let the null
tetrad vectors $Z^\mu_a\,=\,(l^\mu,n^\mu,m^\mu,\bar{m}^\mu)$, with
$a\,=\,1,2,3,4$, undergo the transformation

\begin{equation}Z^\mu_a\rightarrow \tilde{Z}^\mu_a(\tilde{x}^\sigma,\bar{\tilde{x}}^\sigma)\,=\,Z^\rho_a\frac{\partial\tilde{x}
^\mu}{\partial x^\rho}.\end{equation} Obviously, one has to recover the old tetrads and metric as soon as
$\tilde{x}^\sigma\,=\,\bar{\tilde{x}}^\sigma$. In summary, the
effect of the "\emph{tilde transformation}" (\ref{transfo}) is to
generate a new metric whose components are (real) functions of
complex variables, that is

\begin{equation}g_{\mu\nu}\rightarrow \tilde{g}_{\mu\nu}\,:\,\tilde{\mathbf{x}}\times\tilde{\mathbf{x}}\mapsto
\mathbb{R}\end{equation} with

\begin{equation}\tilde{Z}^\mu_a(\tilde{x}^\sigma,\bar{\tilde{x}}^\sigma)|_{\mathbf{x}=\tilde{\mathbf{x}}}
=Z^\mu_a(x^\sigma).\end{equation} For our aims,  we can make
 the choice

\begin{equation}\label{transfo_1}
\tilde{x}^\mu\,=\,x^\mu+ia(\delta^\mu_1-\delta^\mu_0)\cos\theta\rightarrow
\left\{\begin{array}{ll}\tilde{u}\,=\,u+ia\cos\theta\\\\\tilde{r}\,=\,r-ia\cos\theta\\\\
\tilde{\theta}\,=\,\theta\\\\
\tilde{\phi}\,=\,\phi\\\\
\end{array}\right.\end{equation}
where $a$ is constant and the tetrad null vectors
(\ref{tetradvectors}), if we choose
$\tilde{r}\,=\,\bar{\tilde{r}}$, become

\begin{equation}\label{tetradvectors_2}
\left\{\begin{array}{ll}\tilde{l}^\mu\,=\,\delta^\mu_1\\\\\tilde{n}^\mu\,=\,-\frac12
e^{-2\lambda(\tilde{r},\theta)}\delta^\mu_1+
e^{-\lambda(\tilde{r},\theta)-\phi(\tilde{r},\theta)}\delta^\mu_0\\\\
\tilde{m}^\mu\,=\,\frac{1}{\sqrt{2}(\tilde{r}-ia\cos\theta)}\biggl[ia(\delta^\mu_0-\delta^\mu_1)\sin\theta+\delta^\mu_2
+\frac{i}{\sin{\theta}}\delta^\mu_3\biggr]\\\\
\bar{\tilde{m}}^\mu\,=\,\frac{1}{\sqrt{2}(\tilde{r}+ia\cos\theta)}\biggl[-ia(\delta^\mu_0-\delta^\mu_1)\sin\theta+\delta^\mu_2
-\frac{i}{\sin{\theta}}\delta^\mu_3\biggr]\end{array}\right.
\end{equation}

From the transformed null tetrad vectors, a new metric is recovered
using (\ref{eq:gmet}). For the null tetrad vectors given by
(\ref{tetradvectors_2}) and the transformation given by
(\ref{transfo_1}), the new metric, with coordinates
$\tilde{x}^\mu\,=\,(\tilde{u},\tilde{r},\theta,\phi)$, is

\begin{equation}
\tilde{g}^{\mu\nu} = \left(
\begin{array}{cccc} \label{eqn:newmetric}
-\frac{a^2 \sin^2{\theta}}{\Sigma^2} &
e^{-\lambda(\tilde{r},\theta)-\phi(\tilde{r},\theta)} +
\frac{a^2\sin^2{\theta}}{\Sigma^2} & 0 & -\frac{a}{\Sigma^2} \\
. & - e^{-2\lambda(\tilde{r},\theta)} -
\frac{a^2\sin^2{\theta}}{\Sigma^2}
& 0 & \frac{a}{\Sigma^2} \\
. & . & -\frac{1}{\Sigma^2} & 0 \\
. & . & . & -\frac{1}{\Sigma^2\sin^2{\theta}} \\
\end{array}      \right)
\end{equation}
where $\Sigma = \sqrt{\tilde{r}^2 + a^2\cos^2{\theta}}$. In the
covariant form,  the metric (\ref{eqn:newmetric}) is

\begin{eqnarray} \label{eqn:coform}
ds^2&=& e^{2\phi(\tilde{r},\theta)}d\tilde{u}^2
+2 e^{\lambda(\tilde{r},\theta)}d\tilde{u}d\tilde{r}
+2ae^{\phi(\tilde{r},\theta)}[e^{\lambda(\tilde{r},\theta)}-
e^{\phi(\tilde{r},\theta)}] \sin^2{\theta}d\tilde{u}d\phi+\nonumber\\&&
-2a e^{\phi(\tilde{r},\theta)+\lambda(\tilde{r},\theta)}\sin^2{\theta}d\tilde{r}^2d\phi-Sigma^2d\theta^2+\nonumber\\&&
- [\Sigma^2 +
a^2\sin^2{\theta}e^{\phi(\tilde{r},\theta)}(2e^{\lambda(\tilde{r},\theta)}-
e^{\phi(\tilde{r},\theta)})]\sin^2{\theta}d\phi^2
\end{eqnarray}

Since the metric is symmetric, the dots in the matrix  are used to indicate
$g^{\mu\nu} = g^{\nu\mu}$. The form of this metric gives the
general result of the Newman-Janis algorithm starting from  any spherically
symmetric "seed" metric.

The metric given in Eq.~(\ref{eqn:coform}) can be  simplified by a further gauge  transformation so that the only off-diagonal
component is $g_{\phi t}$. This procedure makes it easier to compare with
the standard Boyer-Lindquist form of the Kerr metric
\cite{gravitation} and to interpret physical properties such as the frame
dragging. The coordinates $\tilde{u}$ and $\phi$ can be
redefined in such a way that the metric in the new coordinate
system has the properties described above. More explicitly, if we
define the coordinates in the following way

\begin{equation}d\tilde{u}\,=
\,dt+g(\tilde{r})d\tilde{r}\,\,\,\,\,\text{and}\,\,\,\,\,d\phi\,=\,d\phi+h(\tilde{r})d\tilde{r}\end{equation}
where

\begin{equation}
\left\{\begin{array}{ll}g(\tilde{r})=-\frac{e^{\lambda(\tilde{r},\theta)}
(\Sigma^2+a^2\sin^2{\theta}e^{\lambda(\tilde{r},\theta)+\phi(\tilde{r},\theta)})}
{e^{\phi(\tilde{r},\theta)}(\Sigma^2+a^2\sin^2{\theta}e^{2\lambda(\tilde{r},\theta)})}\\\\h(\tilde{r})
=-\frac{ae^{2\lambda(\tilde{r},\theta)}}{\Sigma^2+a^2\sin^2{\theta}e^{2\lambda(\tilde{r},\theta)}}\end{array}\right.
\end{equation}
after some algebraic manipulations, one finds that, in
$(t,\tilde{r},\theta,\phi)$ coordinates system, the metric
(\ref{eqn:coform}) becomes

\begin{eqnarray} \label{eq:blform}
ds^2&=&e^{2\phi(\tilde{r},\theta)}dt^2+ a
e^{\phi(\tilde{r},\theta)}[e^{\lambda(\tilde{r},\theta)}-
e^{\phi(\tilde{r},\theta)}] \sin^2{\theta} dtd\phi-\nonumber\\&&
 \frac{\Sigma^2}{(\Sigma^2 e^{-2\lambda(\tilde{r},\theta)} +
a^2\sin^2{\theta})}d\tilde{{r}}^2 
-\Sigma^2d\theta^2+\nonumber\\&&
-[\Sigma^2 +
a^2\sin^2{\theta}e^{\phi(\tilde{r},\theta)}(2e^{\lambda(\tilde{r},\theta)}-
e^{\phi(\tilde{r},\theta)})]\sin^2{\theta}d\phi^2
\end{eqnarray}

This metric represents the complete family of
metrics that may be obtained by performing the Newman-Janis
algorithm on any static spherically symmetric "seed" metric, written
in Boyer-Lindquist type coordinates. The validity of these
transformations requires the condition $\Sigma^2+a^2\sin^2\theta
e^{2\lambda(\tilde{r},\theta)}\neq 0$, where  $e^{2\lambda(\tilde{r},\theta)}> 0$. Our task is now to show that such an approach can be used to derive axially symmetric solutions also in $f(R)$-gravity that, possibly, can be regarded as black hole solutions..

\section{Axially symmetric solutions in $f(R)$-gravity}
\label{cinque}

Starting from the above spherically symmetric solution (\ref{sol_noe_2}),
 the metric tensor, written in the Eddington--Finkelstein
 coordinates $(u,r,\theta,\phi)$
of the form (\ref{metrictensorcontro}) is

\begin{equation}\label{metrictensorcontro_noe}
g^{\mu\nu} = \left(
\begin{array}{cccc}
0 & \sqrt{\frac{2}{\beta r}} & 0 & 0     \\
. & -2-\frac{2\alpha}{\beta r}& 0 & 0 \\
. & . & -1/r^2 & 0 \\
. & . & . & -1/(r^2\sin^2{\theta})
\end{array}    \right).
\end{equation}
The complex  tetrad  null vectors (\ref{tetradvectors}) are now

\begin{equation}\label{tetradvectors_noe}
\left\{\begin{array}{ll}l^\mu\,=\,\delta^\mu_1\\\\n^\mu\,=\,-
\biggl[1+\frac{\alpha}{\beta}\biggl(\frac{1}{\bar{r}}+\frac{1}{r}\biggr)\biggr]\delta^\mu_1+
\sqrt{\frac{2}{\beta}}\frac{1}{\sqrt[4]{\bar{r}r}}\delta^\mu_0\\\\
m^\mu\,=\,\frac{1}{\sqrt{2}\bar{r}}(\delta^\mu_2
+\frac{i}{\sin{\theta}}\delta^\mu_3)\end{array}\,.\right.
\end{equation}

By computing the complex coordinates transformation
(\ref{transfo_1}),  the tetrad null vectors become

\begin{equation}\label{tetradvectors_noe_1}
\left\{\begin{array}{ll}\tilde{l}^\mu\,=\,\delta^\mu_1\\\\\tilde{n}^\mu\,=\,-
\biggl[1+\frac{\alpha}{\beta}\frac{\text{Re}\{\tilde{r}\}}{\Sigma^2}\biggr]\delta^\mu_1+
\sqrt{\frac{2}{\beta}}\frac{1}{\sqrt{\Sigma}}\delta^\mu_0\\\\
\tilde{m}^\mu\,=\,\frac{1}{\sqrt{2}(\tilde{r}+ia\cos\theta)}\biggl[ia(\delta^\mu_0-\delta^\mu_1)\sin\theta+\delta^\mu_2
+\frac{i}{\sin{\theta}}\delta^\mu_3\biggr]\end{array}\right.
\end{equation}
Now by performing the same procedure as in previous section,  we derive an axially
symmetric metric of the form (\ref{eq:blform}) but starting from  the spherically symmetric
metric (\ref{sol_noe_2}), that is

\begin{eqnarray}\label{rotating}
ds^2&=&\frac{r(\alpha+\beta r)+a^2\beta\cos^2\theta}{\Sigma}du^2+2 \frac{a(-2\alpha r-2\beta\Sigma^2+\sqrt{2\beta}\Sigma^{3/2})\sin^2\theta}{2\Sigma}dud\phi+\nonumber\\&&
 -\frac{\beta\Sigma^2}{2\alpha r+\beta(a^2+r^2+\Sigma^2)}dr^2-\Sigma^2d\theta^2+\nonumber\\&&
  -\biggl[\Sigma^2 -\frac{a^2(\alpha
r+\beta\Sigma^2-\sqrt{2\beta}\Sigma^{3/2})\sin^2\theta}{\Sigma}\biggr]
\sin^2\theta d\phi^2
\end{eqnarray}

It  is worth noticing that the condition $a=0$ immediately gives  the metric (\ref{sol_noe_2}).  This is nothing else but an example: the method is general and can be extended to any spherically symmetric solution derived in $f(R)$-gravity.

\subsection{Physical applications: geodesics and orbits}
\label{cinque.1}
Let us  discuss now possible physical applications of the above results. We will take into account a  freely falling particle moving in the space-time described by the metric  (\ref{rotating}).  For our aims, we
 make explicit use of the Hamiltonian formalism. Given a metric $g_{\mu\nu}$, the motion along
the geodesics   is described by the Lagrangian
\begin{equation}
{\cal L}(x^{\mu},\dot{x}^\mu)=\dfrac{1}{2}g_{\mu\nu}\dot{x}^{\mu}\dot{x^\nu} \ ,
\end{equation}
where the overdot stands for derivative with respect to an affine parameter
$\lambda$ used to parametrize the curve.  The Hamiltonian description is achieved by considering
the canonical momenta and the Hamiltonian function
\begin{equation}
p_{\mu} = \frac{\de{\cal L}}{\de \dot{x}^{\mu}}=g^{\mu\nu}p_\mu p_\nu \ ,
\;\;\;\;\;\;\;\;\;\;
{\cal H }= p_{\mu}\dot{x}^{\mu}-{\cal L} \ ,
\end{equation}
that results
${\displaystyle
{\cal H}= \frac{1}{2}p_\mu p_\nu g^{\mu\nu} }$.
The advantage of the Hamiltonian formalism with respect to the Lagrangian one is that the resulting
equations of motion do not contain any sign ambiguity coming from turning points  in the orbits (see, for example, \cite{chandra}) .
The Hamiltonian results explicitly independent of time and it is
${\displaystyle
{\cal H}= -\frac{1}{2}m^2 ,}$
where the rest mass $m$ is a constant ($m=0$ for photons).
The geodesic equations are
\begin{equation}
\frac{dx^{\mu}}{d\lambda}=\frac{\partial\cal{ H}}{\partial p_{\mu}}=g^{\mu\nu}p_{\nu}=p^{\mu},
\label{2a}
\end{equation}
\begin{equation}
\frac{dp_{\mu}}{d\lambda}=-\frac{\partial\cal{H}}{\partial x{\mu}}=-\frac{1}{2}\frac{\partial g^{\alpha\beta}}{\partial x^{\mu}}p_{\alpha}p_{\beta}=g^{\gamma\beta}\Gamma^{\alpha}_{\mu\gamma}p_{\alpha}p_{\beta}.
\label{2b}
\end{equation}
In addition, since the Hamiltonian is independent of the affine parameter $\lambda$,
one can directly use the coordinate time as integration parameter.
The problem is so reduced to  solve  six equations of motion.  Using the above definitions, it is easy to achieve the reduced
Hamiltonian (now linear in the momenta)

\begin{equation}
H=-p_{0}=\left[\frac{p_{i}g^{0i}}{g^{00}}+\left[\left(\frac{p_{i}g^{0i}}{g^{00}}\right)^{2}-\frac{m^{2}+p_{i}p_{j}g^{ij}}{g^{00}}\right]^{1/2}\right]\label{eq:e}\end{equation}
with the equations of motion

\begin{equation}
\frac{dx^{i}}{dt}=\frac{\partial H}{\partial p_{i}}\ ,\;\;\;\;\;\;\;\;\;\;\;\;
\frac{dp_{i}}{dt}=-\frac{\partial H}{\partial x^{i}}\ ,\label{eq:4b}\end{equation}
that give the orbits. The method can be applied to the above solution (\ref{rotating}) 
 by which the elements of the inverse metric can be easily obtained:

\begin{eqnarray}
g^{tt}&=&\left\{4 \Sigma ^2 \left[\Sigma ^2-\frac{a^2 \sin ^2\theta \left(r
   \alpha -\sqrt{2} \sqrt{\beta } \Sigma ^{3/2}+\beta  \Sigma
   ^2\right)}{\Sigma }\right]\right\}\times
   \nonumber\\&&
   \left\{a^2 \sin ^2\theta \left(2 r \alpha
   -\sqrt{2} \sqrt{\beta } \Sigma ^{3/2}+2 \beta  \Sigma ^2\right)^2
  +4 \Sigma  \left[a^2 \beta  \cos ^2\theta+r (r \beta +\alpha )\right]\right. \times\nonumber\\&&
  \left. \left( \Sigma ^2-\frac{a^2 \sin ^2\theta \left(r \alpha -\sqrt{2} \sqrt{\beta }\Sigma ^{3/2}+\beta  \Sigma ^2\right)}{\Sigma }\right)\right \}^{-1}\,,
  \nonumber\\
   g^{rr}&=&-\frac{\beta  \left(a^2+r^2+\Sigma ^2\right)+2 r \alpha }{\beta  \Sigma
   ^2}\,,\nonumber\\
 g^{\theta\theta}&=&-\frac{1}{\Sigma ^2}\,,\nonumber\\
 g^{t\phi}&=& \left\{2 a \Sigma  \left(-2 r \alpha +\sqrt{2} \sqrt{\beta } \Sigma
   ^{3/2}-2 \beta  \Sigma ^2\right)\right\}\times\nonumber\\&&
    \left\{a^2 \sin ^2\theta \left(2 r \alpha
   -\sqrt{2} \sqrt{\beta } \Sigma ^{3/2}+2 \beta  \Sigma ^2\right)^2
  +4 \Sigma  \left[a^2 \beta  \cos ^2\theta+r (r \beta +\alpha )\right]\right. \times\nonumber\\&&
  \left. \left( \Sigma ^2-\frac{a^2 \sin ^2\theta \left(r \alpha -\sqrt{2} \sqrt{\beta }\Sigma ^{3/2}+\beta  \Sigma ^2\right)}{\Sigma }\right)\right \}^{-1}\,,
  \nonumber\\
     g^{\phi\phi}&=&-\left\{4 \Sigma  \csc ^2\theta \left[a^2 \beta  \cos ^2\theta +r (r
   \beta +\alpha )\right]\right\}\times\nonumber\\&&
    \left\{a^2 \sin ^2\theta \left(2 r \alpha
   -\sqrt{2} \sqrt{\beta } \Sigma ^{3/2}+2 \beta  \Sigma ^2\right)^2
  +4 \Sigma  \left[a^2 \beta  \cos ^2\theta+r (r \beta +\alpha )\right]\right. \times\nonumber\\&&
  \left. \left( \Sigma ^2-\frac{a^2 \sin ^2\theta \left(r \alpha -\sqrt{2} \sqrt{\beta }\Sigma ^{3/2}+\beta  \Sigma ^2\right)}{\Sigma }\right)\right \}^{-1}\,,
 \end{eqnarray}

and the null ones
  \begin{equation}
g^{tr}=g^{t\theta}=  g^{r\theta}=   g^{r\phi}= g^{\theta\phi}=0\,.
   \end{equation}

\begin{figure}
\begin{tabular}{|c|c|}
\hline
\includegraphics[scale = 0.55]{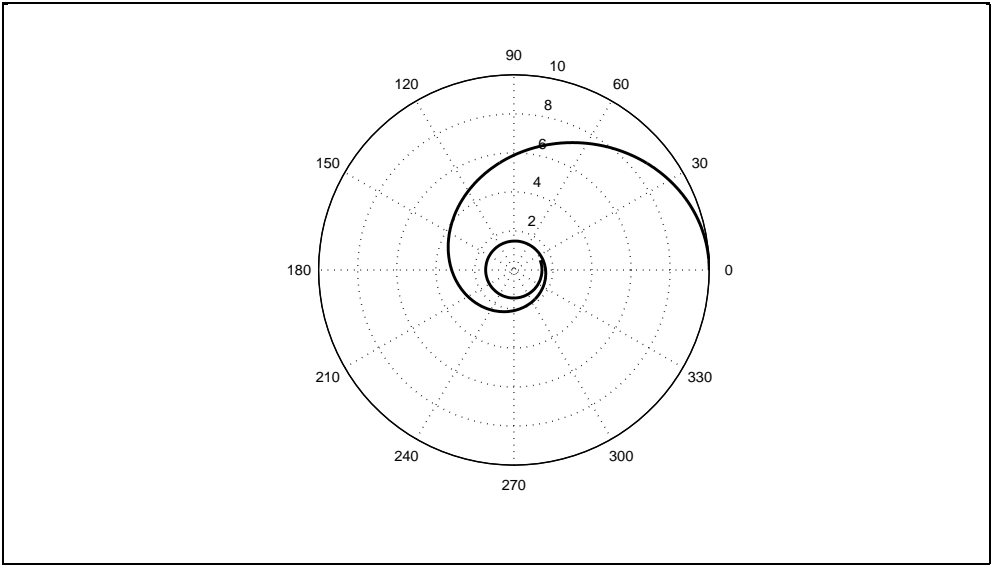}
\tabularnewline
\hline
\end{tabular}
\caption{Relative motion of the test particle with $m=1$.}\label{1}
\end{figure}

Let us consider the equatorial plane, {\it i.e.} $\theta=\frac{\pi}{2}$, $\dot{\theta}=0$, and assume $\alpha=1$ and $\beta=2$.
The reduced  Hamiltonian $H(r,\theta, \phi,p_{r},p_{\theta},p_{\phi};t)=H$ can be written as

\begin{eqnarray}
H&=&\frac{2 a p_{\phi} \left(-2 r^3+r^2-1\right)}{a^2 \left(-2 (r-1) r^2-1\right)+r^5}+
\left\{\left[ \left(4 a^2 p_{\phi}^{2} \left(-2 r^3+r^2-1\right)^2\right.\right.\right.+\nonumber\\ &&
\left.\left.\left.\left.-a^2 \left(-2 (r-1) r^2-1\right)-r^5\right)\left(a^2 \left(r^2 (r (2 r-3) (2 r+1)+6)-2\right)+\right.\right.\right.\right.\nonumber\\&&\left.\left.\left.+(2 r+1) r^4\right)\times\right.\right.\nonumber\\&&\left.\left. \left(-\frac{p_{\phi} (2 r+1)}{a^2 \left(r^2 (r (2 r-3) (2 r+1)+6)-2\right)+(2 r+1) r^4}+\right.\right.\right.\nonumber\\&&\left.\left.\left.-\frac{p_{r} \left(a^2+r^2+r\right)+p_{\theta}}{r^4}-p_{r}+1\right)
\right]\right\}^\frac{1}{2} .\nonumber\\
  \label{eq:8}
   \end{eqnarray}
It is independent of $\phi$ ({\it i.e.} we are considering an azimuthally
symmetric spacetime), and then  the conjugate momentum $p_{\phi}$  is
an  integral of motion.  From Eqs. (\ref{eq:4b}),  one can derive the
coupled equations for $\{r,\theta,\phi,p_{r}$,  $p_{\theta}\}$ and integrate them numerically (the expressions are very cumbersome and will not be reported here). To this goal, we have to
specify the initial value of the position-momentum vector in the phase space.  A  Runge-Kutta method can be used to solve the differential equations. In Fig \ref{1},  the relative trajectories are sketched.

\section{Hydrostatic equilibrium of stellar structures}
\label{sei}

We have shown that it is possible  to derive black hole spherically and axially symmetric solutions  in
$f(R)$-gravity. The physics of black holes contains a lot of unsolved  problems. One of them is the question on the nature of the black
holes as well as the nature of l  stellar objects presenting anomalous behaviors (e.g. the magnetars)  that do not obey the standard dynamics  of (relativistic and non-relativistic) stellar structures. 
For this purpose, it is necessary to understand the hydrostatic equilibrium of such objects, that  
 is one of fundamental properties (or one of the basic assumptions) of self-gravitating systems.  
 Here we start by describing the standard hydrostatic equilibrium assuming Newtonian (i.e. the GR weak field limit) and then we compare these results with the equilibrium derived from the weak field limit of  $f(R)$-gravity.
 
The  condition of hydrostatic equilibrium for  stellar structures  in Newtonian dynamics is achieved by   considering the  equation

\begin{equation}\label{19.1}
\frac{dp}{dr}\,=\,\frac{d\Phi}{dr}\rho\,,
\end{equation}
where $p$ is the pressure, $-\Phi$ is the gravitational potential, and $\rho$ is the density \cite{kippe}.   Together with  the above equation, the Poisson equation

\begin{equation}\label{19.2}
\frac{1}{r^2}\frac{d}{dr}\left(r^2\frac{d\Phi}{dr}\right)\,=\,-4\pi G\rho\,,
\end{equation}
gives the gravitational potential as  solution for a given matter density $\rho$. 
Since we are taking into account only static and stationary situations, here  we consider only time-independent solutions \footnote{The radius $r$  is assumed as the spatial coordinate. It  varies from $r\,=\,0$ at the center to  $r\,=\,\xi$ at the surface of the star}.
In general, the temperature $\tau$ appears in Eqs. (\ref{19.1}) and (\ref{19.2})  the density satisfies an equation of state of the form $\rho\,=\,\rho(p,\tau)$. In any case, we assume that there exists a polytropic relation between $p$ and $\rho$ of the form

\begin{equation}\label{19.3}
p\,=\,K\rho^\gamma\,,
\end{equation}
where $K$ and $\gamma$ are constant. Note that  $\Phi\,>\,0$ in the interior of the model  since we define the gravitational potential as $-\Phi$. The polytropic constant $K$ is fixed and can be obtained as a combination of fundamental  constants. However there are several realistic  cases where $K$ is not fixed and another equation for its evolution is needed.  The constant $\gamma$  is the {\it polytropic exponent }. Inserting  the polytropic equation of state into  Eq. (\ref{19.1}), we obtain
\begin{equation}\label{19.6}
\frac{d\Phi}{dr}\,=\,\gamma K \rho^{\gamma-2}\frac{d\rho}{dr}\,.
\end{equation}
For  $\gamma\neq1$,  the above equation can be integrated giving

\begin{eqnarray}\label{densita}
\frac{\gamma K}{\gamma-1}\rho^{\gamma-1}\,=\,\Phi\,\,\,\,\rightarrow\,\,\,\,\rho\,=\,\biggl[\frac{\gamma-1}{\gamma K}\biggr]^{\frac{1}{\gamma-1}}\Phi^{\frac{1}{\gamma-1}}\,\doteq\,A_n\Phi^n
\end{eqnarray}
where we have chosen the integration constant to give $\Phi=0$ at surface $(\rho=0$). The constant $n$ is called the {\it polytropic index} and is defined as  $n=\frac{1}{\gamma-1}$. Inserting the relation (\ref{densita})  into the Poisson equation, we obtain a differential equation for the gravitational potential
\begin{equation}\label{19.8}
\frac{d^2\Phi}{dr^2}+\frac{2}{r}\frac{d\Phi}{dr}\,=\,-4\pi G A_n\Phi^n\,.
\end{equation}
Let us define now the dimensionless variables

\begin{eqnarray}\label{trans1}
\left\{\begin{array}{ll}
z\,=\,|\mathbf{x}|\sqrt{\frac{\mathcal{X}A_n\Phi_c^{n-1}}{2}}\\\\
w(z)\,=\,\frac{\Phi}{\Phi_c}\,=\,(\frac{\rho}{\rho_c})^\frac{1}{n}
\end{array}\right.
\end{eqnarray}
where the subscript $c$ refers to the center of the star and  the relation between $\rho$ and $\Phi$ is given by Eq. (\ref{densita}). At the center  $(r\,=\,0)$,  we have $z\,=\,0$, $\Phi\,=\,\Phi_c$, $\rho\,=\,\rho_c$ and therefore $w\,=\,1$. Then 
Eq. (\ref{19.8}) can be written

\begin{eqnarray}\label{LE}
\frac{d^2w}{dz^2}+\frac{2}{z}\frac{dw}{dz}+w^n\,=\,0
\end{eqnarray}
This is the standard {\it Lane-Embden equation} describing the hydrostatic equilibrium of stellar structures in the Newtonian theory \cite{kippe}. Now we want to compare this standard result with the one coming from $f(R)$-gravity.

\section{The Newtonian limit of $f(R)$-gravity}\label{sette}
In order to achieve the Newtonian limit of the theory
 the metric tensor $g_{\mu\nu}$  have to be approximated as follows 

\begin{eqnarray}\label{metric_tensor_PPN}
  g_{\mu\nu}\,\sim\,\begin{pmatrix}
  1-2\,\Phi(t,\mathbf{x})+\mathcal{O}(4)& \mathcal{O}(3) \\
  \\
  \mathcal{O}(3)& -\delta_{ij}+\mathcal{O}(2)\end{pmatrix}
\end{eqnarray}
where $\mathcal{O}(n)$ (with $n\,=$ integer) denotes the order of the expansion (see \cite{mio} for details).
The set of coordinates\footnote{The Greek index runs between $0$
and $3$; the Latin index between $1$ and $3$.} adopted is
$x^\mu\,=\,(t,x^1,x^2,x^3)$. The Ricci scalar formally becomes

\begin{eqnarray}
R\,\sim\,R^{(2)}(t,\mathbf{x})+\mathcal{O}(4)
\end{eqnarray}
The $n$-th derivative of Ricci function can be developed as

\begin{eqnarray}
f^{n}(R)\,\sim\,f^{n}(R^{(2)}+\mathcal{O}(4))\,\sim\,f^{n}(0)+f^{n+1}(0)R^{(2)}+\mathcal{O}(4)
\end{eqnarray}
here $R^{(n)}$ denotes a quantity of order  $\mathcal{O}(n)$.
From lowest order of field equations, we
have $f(0)\,=\,0$ which trivially follows from the above assumption (\ref{metric_tensor_PPN}) for the metric.  This means that  the space-time
is asymptotically Minkowskian and we are discarding a cosmological constant term in this analysis\footnote{This assumption is quite natural since the contribution of a cosmological constant term is irrelevant at stellar level. }. Field eqautaions  at $\mathcal{O}(2)$-order, that is at Newtonian level, are

\begin{eqnarray}\label{PPN-field-equation-general-theory-fR-O2}
\left\{\begin{array}{ll}
R^{(2)}_{tt}-\frac{R^{(2)}}{2}-f''(0)\triangle
R^{(2)}\,=\,\mathcal{X}\,T^{(0)}_{tt}
\\\\
-3f''(0)\triangle
R^{(2)}-R^{(2)}\,=\,\mathcal{X}\,T^{(0)}
\end{array}\right.
\end{eqnarray}
where $\triangle$ is the Laplacian in the flat space, $R^{(2)}_{tt}\,=\,-\triangle\Phi(t,\mathbf{x})$ and, for the sake of  simplicity, we set $f'(0)\,=\,1$. We recall that the energy-momentum tensor  for a perfect fluid  is

\begin{eqnarray}
T_{\mu\nu}\,=\,(\epsilon+p)\,u_{\mu} u_{\nu}-p\,g_{\mu\nu}
\end{eqnarray}
where $p$ is the pressure  and $\epsilon$ is the energy density. Being the pressure contribution negligible in the field equations in Newtonian approximation,  we have

\begin{eqnarray}\label{HOEQ}
\left\{\begin{array}{ll}
\triangle\Phi+\frac{R^{(2)}}{2}+f''(0)\triangle R^{(2)}\,= \,-\mathcal{X}\rho
\\\\
3f''(0)\triangle R^{(2)}+R^{(2)}\,=\,-\mathcal{X}\rho
\end{array}\right.
\end{eqnarray}
where $\rho$ is now the mass density\footnote{Generally it is $\epsilon\,=\,\rho\,c^2$.}. We note that  for $f''(0)\,=\,0$ we have the  standard Poisson equation: $\triangle\Phi\,=\,-4\pi G\rho$. This means that as soon as the second derivative of $f(R)$ is different from zero, deviations from the Newtonian limit of GR emerge.

The gravitational potential $-\Phi$, solution of Eqs. (\ref{HOEQ}), has in general
a Yukawa-like behavior depending on a
characteristic length on which it evolves \cite{mio}. Then as it is evident
the Gauss theorem is not valid\footnote{It is worth noticing that also if the Gauss theorem does not hold, the Bianchi identities are always valid so the conservation laws are guaranteed. } since the force
law is not $\propto|\mathbf{x}|^{-2}$. The equivalence between a
spherically symmetric distribution and point-like distribution is
not valid and how the matter is distributed in the space is very
important \cite{mio,stabile,cqg}.

Besides the Birkhoff theorem results modified at Newtonian level:
the solution can be only factorized by a space-depending function
and an arbitrary time-depending function \cite{mio}. Furthermore
the correction to the gravitational potential is depending on
the only first two derivatives of $f(R)$ in $R\,=\,0$. This means that different analytical 
theories, from the third derivative perturbation terms on,  admit the same Newtonian
limit  \cite{mio,stabile}.

 Eqs. (\ref{HOEQ}) can be considered the \emph{modified Poisson equation}
for $f(R)$-gravity. They  do not depend on gauge condition choice \cite{cqg}.
We know that
\begin{eqnarray}
R^{(2)}&\simeq \frac{1}{2}\nabla^2 g^{(2)}_{00}- \frac{1}{2}\nabla^2g^{(2)}_{ii}\,.\label{3}
\end{eqnarray}
Inserting in the above result the $g_{\mu\nu}$ approximations \eqref{metric_tensor_PPN} we obtain
\begin{eqnarray}
{R^{(2)}} \simeq {\nabla ^2}(\Phi  - \Psi ).
\end{eqnarray}
Finally, we obtain the field equations
\begin{eqnarray}\label{EQP}
&& \nabla^2\Phi+\nabla^2\Psi -2 f''(0) \nabla^4 \Phi + 2 f''(0) \nabla^4 \Psi=2\mathcal{X} \rho \\
\nonumber \\
&& \nabla^2\Phi-\nabla^2\Psi +3 f''(0)\nabla^4\Phi-3 f''(0) \nabla^4 \Psi=-\mathcal{X} \rho \label{EQP2}\,.
\end{eqnarray}
By eliminating the higher-order terms,  the standard Poisson equation is recovered.
This last step, will be useful to calculate the Jeans instability.

\section{Stellar Hydrostatic Equilibrium in $f(R)$-gravity}\label{otto}

From the Bianchi identity,  satisfied by the field equations,  we have

\begin{eqnarray}\label{equidr}
{T^{\mu\nu}}_{;\mu}\,=\,0\,\,\,\,\rightarrow\,\,\,\,\frac{\partial p}{\partial x^k}\,=\,-\frac{1}{2}(p+\epsilon)\frac{\partial \ln g_{tt}}{\partial x^k}
\end{eqnarray}
If the dependence on the temperature  $\tau$ is negligible, \emph{i.e.} $\rho\,=\,\rho(p)$,  this relation can be introduced into Eqs. (\ref{HOEQ}), which become a system of three equations for $p$, $\Phi$ and $R^{(2)}$ and can be solved without the other structure equations.

Let us suppose that matter satisfies still a polytropic equation $p\,=\,K\,\rho^\gamma$. If we introduce Eq.(\ref{densita}) into Eqs. (\ref{HOEQ}) we obtain an integro-differential equation for the gravitational potential $-\Phi$, that is 

\begin{eqnarray}\label{deltafi}
\triangle\Phi(\mathbf{x})+\frac{2\mathcal{X}A_n}{3}\Phi(\mathbf{x})^n\,=\,-\frac{m^2\mathcal{X}A_n}{6} \int d^3\mathbf{x}'\mathcal{G}(\mathbf{x},\mathbf{x}')\Phi(\mathbf{x}')^n
\end{eqnarray}
where ${\displaystyle \mathcal{G}(\mathbf{x},\mathbf{x}')\,=\,-\frac{1}{4\pi}\frac{e^{-m|\mathbf{x}-\mathbf{x}'|}}{|\mathbf{x}-\mathbf{x}'|}}$
is the Green function of  the  field operator $\triangle_\mathbf{x}-m^2$ for systems with spherical symmetry and ${\displaystyle m^2\,=\,-\frac{1}{3f''(0)}}$ (for details see \cite{stabile,cqg}). The integro-differential nature of Eq.(\ref{deltafi}) is the proof of the non-viability of Gauss theorem for $f(R)$-gravity. Adopting again the dimensionless variables

\begin{eqnarray}\label{trans}
\left\{\begin{array}{ll}
z\,=\,\frac{|\mathbf{x}|}{\xi_0}\\\\
w(z)\,=\,\frac{\Phi}{\Phi_c}
\end{array}\right.
\end{eqnarray}
where

\begin{eqnarray}\label{charpar}
\xi_0\,\doteq\,\sqrt{\frac{3}{2\mathcal{X}A_n\Phi_c^{n-1}}}
\end{eqnarray}
is a characteristic length linked to stellar radius $\xi$, Eq. (\ref{deltafi}) becomes

\begin{eqnarray}\label{LEmod}
\frac{d^2w(z)}{dz^2}+\frac{2}{z}\frac{d w(z)}{dz}+w(z)^n\,=\,\frac{m\xi_0}{8}\frac{1}{z}\int_0^{\xi/\xi_0}
dz'\,z'\,\biggl\{e^{-m\xi_0|z-z'|}-e^{-m\xi_0|z+z'|}\biggr\}\,w(z')^n\nonumber\\
\end{eqnarray}
which is the \emph{modified Lan\'{e}-Emden equation} deduced from $f(R)$-gravity. Clearly  the particular $f(R)$-model is specified by the parameters $m$ and $\xi_0$. 
If $m\,\rightarrow\,\infty$ (\emph{i.e.} $f(R)\,\rightarrow\,R$),  Eq. (\ref{LEmod}) becomes Eq. (\ref{LE}). We are only interested in solutions of Eq. (\ref{LEmod}) that are finite at the center, that is for $z\,=\,0$. Since the center must be an equilibrium point,  the gravitational acceleration $|\mathbf{g}|\,=\,-d\Phi/dr\,\propto\,dw/dz$ must vanish for $w'(0)\,=\,0$. Let us assume we have solutions $w(z)$ of Eq.(\ref{LEmod}) that fulfill the  boundary conditions $w(0)\,=\,1$ and $w(\xi/\xi_0)\,=\,0$; then according to the choice (\ref{trans}), the radial distribution of  density is given by

\begin{eqnarray}
\rho(|\mathbf{x}|)\,=\,\rho_cw^n\,,\,\,\,\,\,\,\,\,\rho_c\,=\,A_n{\Phi_c}^n
\end{eqnarray}
and the pressure by

\begin{eqnarray}
p(|\mathbf{x}|)\,=\,p_cw^{n+1}\,,\,\,\,\,\,\,\,\,p_c\,=\,K{\rho_c}^\gamma
\end{eqnarray}

For $\gamma\,=\,1$ (or $n\,=\,\infty$) the integro-differential Eq. (\ref{LEmod}) is not correct. This means that the theory does not contain the case of isothermal sphere of ideal gas. 
In this case, the polytropic relation is $p\,=\,K\,\rho$. Putting this relation into Eq.(\ref{equidr}) we have
\begin{eqnarray}\label{densitaiso}
\frac{\Phi}{K}\,=\,\ln\rho-\ln\rho_c\,\,\,\,\rightarrow\,\,\,\,\rho\,=\,\rho_c\,e^{\Phi/K}
\end{eqnarray}
where  the constant of integration is chosen in such a way that the gravitational potential is zero at the center. If we introduce Eq.(\ref{densitaiso}) into Eqs. (\ref{HOEQ}), we have

\begin{eqnarray}\label{deltafiiso}
\triangle\Phi(\mathbf{x})+\frac{2\mathcal{X}\rho_c}{3}e^{\Phi(\mathbf{x})/K}\,=\,-\frac{m^2\mathcal{X}\rho_c}{6} \int d^3\mathbf{x}'\mathcal{G}(\mathbf{x},\mathbf{x}')e^{\Phi(\mathbf{x}')/K}
\end{eqnarray}
Assuming the dimensionless variables $z\,=\,\frac{|\mathbf{x}|}{\xi_1}$ and $w(z)\,=\,\frac{\Phi}{K}$ where $\xi_1\,\doteq\,\sqrt{\frac{3K}{2\mathcal{X}\rho_c}}$, Eq. (\ref{deltafiiso}) becomes

\begin{eqnarray}\label{LEmodiso}
\frac{d^2w(z)}{dz^2}+\frac{2}{z}\frac{d w(z)}{dz}+e^{w(z)}\,=\,\frac{m\xi_1}{8}\frac{1}{z}\int_0^{\xi/\xi_1}
dz'\,z'\,\biggl\{e^{-m\xi_1|z-z'|}-e^{-m\xi_1|z+z'|}\biggr\}\,e^{w(z')}
\end{eqnarray}
which is the \emph{modified "isothermal" Lan\'{e}-Emden equation} derived $f(R)$-gravity.

\section{ Solutions of the standard and modified   Lan\'{e}-Emden Equations}\label{nove}
\begin{figure}[htbp]
  \centering
  \includegraphics[scale=1]{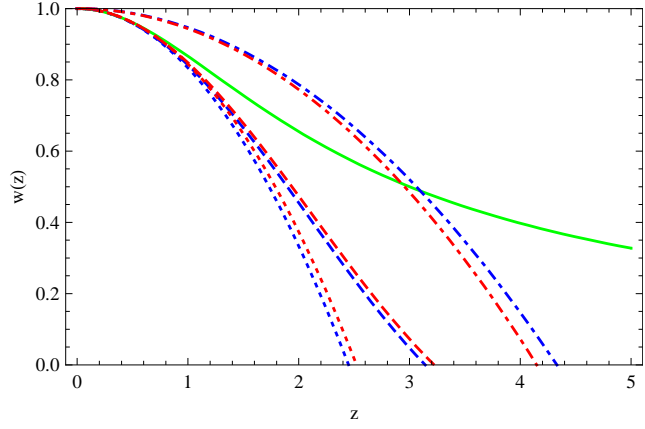}\\
  \caption{Plot of solutions (blue lines) of standard Lan\'{e}-Emden Eq. (\ref{LE}): $w^{(0)}_{GR}(z)$ (dotted line) and $w^{(1)}_{GR}(z)$ (dashed line). The green line corresponds to $w^{(5)}_{GR}(z)$. The red lines are the solutions of modified Lan\'{e}-Emden Eq. (\ref{LEmod}): $w^{(0)}_{_{f(R)}}(z)$ (dotted line) and $w^{(1)}_{_{f(R)}}(z)$ (dashed line). The blue dashed-dotted line is the  potential  derived from GR ($w_{GR}(z)$) and the red dashed-dotted line the  potential derived from  $f(R)$-gravity ($w_{_{f(R)}}(z)$) for a uniform spherically symmetric mass distribution. The assumed values are $m\xi\,=\,1$ and $m\xi_0\,=\,4$. From a rapid inspection of these plots, the differences between GR and $f(R)$ gravitational potentials are clear and the tendency is that at larger radius $z$ they become more evident. }
  \label{fig}
\end{figure}

The task is now to solve the modified Lan\'{e}-Emden equation and compare its solutions to those of standard Newtonian theory.
Only for three values of $n$,  the solutions of Eq.(\ref{LE}) have analytical expressions \cite{kippe}

\begin{eqnarray}\label{LEsol}
&&n\,=\,0\,\,\,\,\rightarrow\,\,\,\,w^{(0)}_{GR}(z)\,=\,1-\frac{z^2}{6}\nonumber\\
&&n\,=\,1\,\,\,\,\rightarrow\,\,\,\,w^{(1)}_{GR}(z)\,=\,\frac{\sin z}{z}\\
&&n\,=\,5\,\,\,\,\rightarrow\,\,\,\,w^{(5)}_{GR}(z)\,=\,\frac{1}{\sqrt{1+\frac{z^2}{3}}}\nonumber
\end{eqnarray}
We label these solution with $_{GR}$ since they agree with the Newtonian limit of GR.
The surface of the polytrope of index $n$ is defined by the value $z\,=\,z^{(n)}$, where $\rho\,=\,0$ and thus $w\,=\,0$. For $n\,=\,0$ and $n\,=\,1$ the surface is reached for a finite value of $z^{(n)}$. The case $n\,=\,5$ yields a model of infinite radius. It can be shown that for $n\,<\,5$ the radius of polytropic models is finite; for $n\,>\,5$ they have infinite radius. From Eqs.(\ref{LEsol}) one finds $z^{(0)}_{GR}\,=\,\sqrt{6}$, $z^{(1)}_{GR}\,=\,\pi$ and $z^{(5)}_{GR}\,=\,\infty$. A general property of the solutions is that $z^{(n)}$ grows monotonically with the polytropic index $n$. In Fig. \ref{fig} we show the behavior of solutions $w^{(n)}_{GR}$ for $n\,=\,0,\,1,\,5$.
Apart from the three cases where analytic solutions are known, the classical Lan\'{e}-Emden Eq. (\ref{LE}) has to be be solved numerically, considering with the expression

\begin{eqnarray}\label{taydev}
w^{(n)}_{GR}(z)\,=\,\sum_{i\,=\,0}^\infty a^{(n)}_iz^i
\end{eqnarray}
for the neighborhood of the center. Inserting Eq.(\ref{taydev}) into Eq. (\ref{LE}) and by comparing coefficients one finds, at lowest orders, a classification of solutions by the index $n$, that is

\begin{eqnarray}\label{gensolGR}
w^{(n)}_{GR}(z)\,=\,1-\frac{z^2}{6}+\frac{n}{120}z^4+\dots
\end{eqnarray}
The case  $\gamma\,=\,5/3$ and $n\,=\,3/2$ is the non-relativistic limit  while the case  $\gamma\,=\,4/3$ and $n\,=\,3$ is the relativistic limit of a completely degenerate gas.

Also for modified Lan\'{e}-Emden Eq. (\ref{LEmod}),  we have an  exact solution for $n\,=\,0$. In fact, it is straightforward to find out 

\begin{eqnarray}\label{LEmodsol0}
w^{(0)}_{_{f(R)}}(z)\,=\,1-\frac{z^2}{8}+\frac{(1+m\xi)e^{-m\xi}}{4m^2{\xi_0}^2}\biggl[1-\frac{\sinh m\xi_0 z}{m\xi_0 z}\biggr]
\end{eqnarray}
where  the boundary conditions $w(0)\,=\,1$ and $w'(0)\,=\,0$ are satisfied. A comment on the GR limit (that is  $f(R)\rightarrow R$) of solution (\ref{LEmodsol0}) is necessary. In fact when we perform the limit $m\,\rightarrow\,\infty$, we do not recover  exactly $w^{(0)}_{GR}(z)$. The difference is in the definition of quantity $\xi_0$. In $f(R)$-gravity we have the definition (\ref{charpar}) while in GR it is ${\displaystyle \xi_0\,=\,\sqrt{\frac{2}{\mathcal{X}A_n\Phi_c^{n-1}}}}$, since in the first equation of (\ref{HOEQ}), when we perform $f(R)\rightarrow R$, we have to eliminate the trace equation condition. In general, this means that the Newtonian limit and the Eddington parameterization of different relativistic theories of gravity cannot coincide with those of GR (see \cite{eddington} for further details on this point).

The point $z_{_{f(R)}}^{(0)}$ is calculated by imposing $w^{(0)}_{_{f(R)}}(z_{_{f(R)}}^{(0)})\,=\,0$ and by considering the Taylor expansion 

\begin{equation}
\frac{\sinh m\xi_0z}{m\xi_0z}\,\sim\,1+\frac{1}{6}(m\xi_0z)^2+\mathcal{O}(m\xi_0z)^4
\end{equation}
we obtain ${\displaystyle z_{_{f(R)}}^{(0)}\,=\,\frac{2\sqrt{6}}{\sqrt{3+(1+m\xi)e^{-m\xi}}}}$.
Since the stellar radius $\xi$ is given by definition $\xi\,=\,\xi_0\,z_{_{f(R)}}^{(0)}$, we obtain the  constraint

\begin{eqnarray}\label{radius_constraint}
\xi\,=\,\sqrt{\frac{3\Phi_c}{2\pi G}}\frac{1}{\sqrt{1+\frac{1+m\xi}{3}e^{-m\xi}}}
\end{eqnarray}
By solving numerically the constraint\footnote{In principle, there is a solution for any value of $m$.} Eq.(\ref{radius_constraint}), we find the modified expression of the radius $\xi$. If $m\,\rightarrow\,\infty$ we have the standard expression $\xi\,=\,\sqrt{\frac{3\Phi_c}{2\pi G}}$ valid for the Newtonian limit of GR. Besides, it is worth noticing  that in the  $f(R)$-gravity case, for $n=0$, the radius is smaller than in GR. On the other hand,  the gravitational potential $-\Phi$ gives rise to a deeper potential well than the corresponding Newtonian potential derived from GR \cite{stabile}. 

In the case $n\,=\,1$, Eq. (\ref{LEmod}) can be recast as follows

\begin{eqnarray}\label{LEmod2}
\frac{d^2\tilde{w}(z)}{dz^2}+\tilde{w}(z)\,=\,\frac{m\xi_0}{8}\int_0^{\xi/\xi_0}
dz'\,\biggl\{e^{-m\xi_0|z-z'|}-e^{-m\xi_0|z+z'|}\biggr\}\,\tilde{w}(z')
\end{eqnarray}
where $\tilde{w}\,=\,z\,w$. If we consider the solution of (\ref{LEmod2}) as a small perturbation to the one of GR, we have

\begin{eqnarray}\label{hypsol}
\tilde{w}^{(1)}_{_{f(R)}}(z)\,\sim\,\tilde{w}^{(1)}_{GR}(z)+e^{-m\xi}\Delta\tilde{w}^{(1)}_{_{f(R)}}(z)
\end{eqnarray}
The coefficient $e^{-m\xi}\,<\,1$ is the parameter with respect to which we perturb Eq. (\ref{LEmod2}). Besides these position  ensure us that when $m\,\rightarrow\,\infty$ the solution converge to something like 
$\tilde{w}^{(1)}_{GR}(z)$. Substituting Eq.(\ref{hypsol}) in Eq.(\ref{LEmod2}),  we have

\begin{eqnarray}
\frac{d^2\Delta\tilde{w}^{(1)}_{_{f(R)}}(z)}{dz^2}+\Delta\tilde{w}^{(1)}_{_{f(R)}}(z)\,=\,\frac{m\xi_0\,e^{m\xi}}{8}\int_0^{\xi/\xi_0}
dz'\,\biggl\{e^{-m\xi_0|z-z'|}-e^{-m\xi_0|z+z'|}\biggr\}\,\tilde{w}^{(1)}_{GR}(z')\nonumber\\
\end{eqnarray}
and the solution is easily found

\begin{eqnarray}
&& w^{(1)}_{_{f(R)}}(z)\sim \frac{\sin z}{z}\biggl\{1+\frac{m^2{\xi_0}^2}{8(1+m^2{\xi_0}^2)}\biggl[1+\frac{2\,e^{-m\xi}}{1+m^2{\xi_0}^2}(\cos\xi/\xi_0+m\xi_0
\sin\xi/\xi_0)
\biggr]\biggr\}
\nonumber\\\nonumber\\&&
-\frac{m^2{\xi_0}^2}{8(1+m^2{\xi_0}^2)}\biggl[\frac{2\,e^{-m\xi}}{1+m^2{\xi_0}^2}(\cos\xi/\xi_0+m\xi_0\sin\xi/\xi_0)
\frac{\sinh m\xi_0z}{m\xi_0z}+\cos z\biggr]\nonumber\\
\end{eqnarray}
Also in this case,  for $m\,\rightarrow\,\infty$, we do not  recover exactly $w^{(1)}_{GR}(z)$. The reason is the same of previous $n\,=\,0$ case \cite{eddington}. Analytical solutions for other values of $n$ are not available.

To conclude this section,  we report  the  gravitational potential profile generated by a spherically symmetric source of  uniform mass  with radius $\xi$.
We can impose a mass density of the form 
 \begin{equation}
 \rho\,=\,\frac{3M}{4\pi\xi^3}\Theta(\xi-|\mathbf{x}|)
 \end{equation}
  where $\Theta$ is the Heaviside function and $M$ is the mass \cite{stabile,cqg}. By solving field Eqs. (\ref{HOEQ}) {\it inside the star}  and considering the boundary conditions $w(0)\,=\,1$ and $w'(0)\,=\,0$, we get
\begin{eqnarray}\label{sol_pot}
w_{_{f(R)}}(z)\,&=&\,\biggl[\frac{3}{2\xi}+\frac{1}{m^2\xi^3}-\frac{e^{-m\xi}(1+m\,\xi)}{m^2\xi^3}\biggr]^{-1}\times
\nonumber\\ && \times\biggl[\frac{3}{2\xi}+\frac{1}{m^2\xi^3}-\frac{{\xi_0}^2z^2}{2\xi^3}-\frac{e^{-m\xi}(1+m\,\xi)}{m^2\xi^3}\frac{\sinh m\xi_0z}{m\xi_0z}\biggr]
\end{eqnarray}
In the limit $m\,\rightarrow\,\infty$, we recover the GR case $w_{GR}(z)\,=\,1-\frac{{\xi_0}^2z^2}{3\xi^2}$. In Fig. \ref{fig} we show the behaviors of $w^{(0)}_{_{f(R)}}(z)$ and $w^{(1)}_{_{f(R)}}(z)$ with respect to the corresponding GR cases. Furthermore, we plot the potential generated by  a uniform spherically symmetric  mass distribution in GR and $f(R)$-gravity and the case $w^{(5)}_{GR}(z)$. 

\section{Examples: dust-dominated self-gravitating systems}
\label{dieci}
From the above equations of hydrostatic equilibrium in $f(R)$-gravity, one can  study the formation and collapse of self gravitating objects like stars and black holes. 
It is  well known that
the  scenario involving the formation of cosmic structures ({\it i.e.} from  stars up to galaxies and clusters of galaxies) occurs for gravitational instability.
 In the standard picture, the first objects that form are the dark matter  halos, which aggregate in a hierarchical way due to gravitational collapse. If such systems  reach a virial equilibrium, the collapse can stop otherwise it can indefinitely continue.
Subsequently, the baryons are affected by the gravity of the halo potential wells: gas, of collisional nature, converts the kinetic energy of the 'fall' into thermal energy and heat reaching the virial temperature. 
 Subsequently, the radiation losses cause the cooling of the baryonic component, its condensation and subsequent formation of molecular clouds, finally of stars.
The standard theory of gravitational collapse for dust dominated systems and can be compared to that for $f(R)$-gravity.
The  collapse of self-gravitational collisionless systems can be dealt with the  introduction  of coupled
collisionless Boltzmann and Poisson equations  (for details, see \cite{BT}):

\begin{equation}\label{Boltz_poisson1}
\dfrac{{\partial f(\vec r,\vec v,t)}}{{\partial t}} + \left( {\vec v\cdot\vec \nabla _r } \right)f(\vec r,\vec v,t)  -\left( {\vec \nabla \Phi \cdot\vec \nabla _v } \right)f(\vec r,\vec v,t) = 0
\end{equation}

\begin{equation}\label{Boltz_poisson2}
 \vec \nabla ^2 \Phi (\vec r,t) = 4\pi G\int {f(\vec r,\vec v,t)} d\vec v.
\end{equation}

A self-gravitating system at equilibrium is described by a time-independent distribution function $f_0(x,v)$ and a potential $\Phi_0(x)$ that are solutions
 of  Eq.\eqref{Boltz_poisson1} and \eqref{Boltz_poisson2}. Considering a small perturbation to this equilibrium:

\begin{eqnarray}
 &&  f(\vec r,\vec v,t) = f_0 (\vec r,\vec v) + \epsilon f_1 (\vec r,\vec v,t),
 \\
 &&  \Phi (\vec r,t) = \Phi _0 (\vec r) + \epsilon \Phi _1 (\vec r,t),
\end{eqnarray}

where $\epsilon \ll 1$ and by substituting in Eq. \eqref{Boltz_poisson1} and \eqref{Boltz_poisson2} and by linearizing, one obtains:
\begin{equation}\label{Linear_Boltz_poisson1}
 \dfrac{{\partial f_1 (\vec r,\vec v,t)}}{{\partial t}} + \vec v\cdot\dfrac{{\partial f_1 (\vec r,\vec v,t)}}{{\partial \vec r}} 
-\vec \nabla \Phi _1 (\vec r,t)\cdot\dfrac{{\partial f_0 (\vec r,\vec v)}}{{\partial \vec v}} - \vec\nabla \Phi _0 (\vec r)\cdot\dfrac{{\partial f_1 (\vec r,\vec v,t)}}{{\partial \vec v}} = 0\,,
\end{equation}
\begin{equation}\label{Linear_Boltz_poisson2}
 \vec \nabla ^2 \Phi _1 (\vec r,t) = 4\pi G\int {f_1 (\vec r,\vec v,t)} d\vec v\,,
\end{equation}

Since the equilibrium state is assumed to be homogeneous and time-independent, one can  set $f_0(\vec x, \vec v,t) = f_0(\vec v)$,
and  the so-called Jeans "swindle" to set $\Phi_0 = 0$. In Fourier components, Eqs.\eqref{Linear_Boltz_poisson1} and
\eqref{Linear_Boltz_poisson2} become:
\begin{eqnarray}
 &&- i\omega f_1  + \vec v\cdot\left( {i\vec kf_1 } \right) - \left( {i\vec k\Phi _1 } \right)\cdot\frac{{\partial f_0 }}{{\partial \vec v}} = 0\,, \\
 &&- k^2 \Phi _1  = 4\pi G\int {f_1 } d\vec v.
 \end{eqnarray}
By combining these equations, the dispersion relation
\begin{equation}\label{eqDISP}
{1 + \frac{{4\pi G}}{{k^2 }}\int {\dfrac{{\vec k\cdot\dfrac{{\partial f_0 }}{{\partial \vec v}}}}{{\vec v\cdot\vec k - \omega }}} d\vec v}=0;
\end{equation}
is obtained.
In the case of stellar systems, by assuming a Maxwellian distribution function for $f_0$, we have
\begin{equation}\label{Maxwellian}
f_0  = \frac{{\rho _0 }}{{(2\pi \sigma ^2 )^{\frac{3}{2}} }}e^{ - \dfrac{{v^2 }}{{2\sigma ^2 }}},
\end{equation}
 imposing that $\vec k=(k,0,0)$ and  substituting in  Eq.\eqref{eqDISP}, one gets:
\begin{equation}
1 - \frac{{2\sqrt {2\pi } G\rho _0 }}{{k\sigma ^3 }}\int {\dfrac{{v_x e^{ - \dfrac{{v_x^2 }}{{2\sigma ^2 }}} }}{{kv_x  - \omega }}} dv_x  = 0.
\end{equation}
By setting $\omega=0$,  the limit for instability is obtained:
\begin{equation}
     k^2(\omega=0)  = \dfrac{{4\pi G\rho _0 }}{{\sigma ^2 }} = k_J^2,
\end{equation}
by which it is possible to  define the Jeans mass ($M_J$) as the mass originally contained within a sphere of diameter $\lambda_J$:
\begin{equation}\label{MJ}
 M_J= \frac{4 \pi}{3} \rho_0 \left(\frac{1}{2} \lambda_J \right)^3,
\end{equation}
where
\begin{equation}
   \lambda_J^2= \dfrac{\pi \sigma^2}{G \rho_0}
   \label{length}
\end{equation}
is the Jeans length.
Substituting Eq. \eqref{length} into  Eq. \eqref{MJ}, we recover
\begin{equation}\label{Mass}
{{{ M}_J} = \dfrac{\pi }{6}\sqrt {\dfrac{1}{{{\rho _0}}}{{\left( {\dfrac{{\pi {\sigma ^2}}}{{G}}} \right)}^3}} }\,.
\end{equation}
All perturbations with wavelengths $\lambda >\lambda_J$  are unstable in the stellar system. In order to evaluate the integral
 in the dispersion relation for real and nonzero values of $\omega$, the dispersion relation has to be rewritten as
\begin{equation}
1 - \dfrac{{k_{_J }^2 }}{{k^2 }}W\left( {\dfrac{\omega}{{k\sigma }}} \right) = 0,
\end{equation}
defining
\begin{equation}
 W\left( {\dfrac{\omega}{{k\sigma }}} \right) \equiv \dfrac{1}{{\sqrt {2\pi } }}\int {\dfrac{{xe^{ - \dfrac{{x^2 }}{2}} }}{{x - Z}}} dx,
\end{equation}

and setting $\omega=i\omega_I$ and $Re(W\left( {\dfrac{\omega}{{k\sigma }}} \right))=0$. In order to study unstable modes (for  details, see 
Appendix B in \cite{BT}) we replace the following identities
\[
\left\{ \begin{array}{l}
 \int\limits_0^\infty  {\dfrac{{x^2 e^{ - x^2 } }}{{x^2  + \beta ^2 }}dx}  = \dfrac{1}{2}\sqrt \pi   - \dfrac{1}{2}\pi \beta e^{\beta ^2 } \left[ {1 - {\rm{erf }}\beta } \right]\,, \\
  \\
 {\rm{erf }}z = \dfrac{2}{{\sqrt \pi  }}\int\limits_0^z {e^{ - t^2 } dt}.  \\
 \end{array} \right.
\]
into the dispersion relation obtaining:
\begin{equation}
\small{k^2  = k_J^2 \left\{ {1 -  \dfrac{{\sqrt \pi \omega _I }}{{\sqrt 2 k\sigma }}e^{\left( {\dfrac{{\omega _I }}{{\sqrt 2 k\sigma }}} \right)^2 } \left[ {1 - {\rm{erf }}\left( {\frac{{\omega _I }}{{\sqrt 2 k\sigma }}} \right)} \right]} \right\}}.
\end{equation}
This is the standard dispersion relation describing the  criterion to collapse for infinite homogeneous fluid and stellar systems \cite{BT}.

\section{Jeans criterion for gravitational instability in $f(R)$-gravity}
\label{undici}
\begin{figure}[!h]
 \centering
  \includegraphics[scale=0.39]{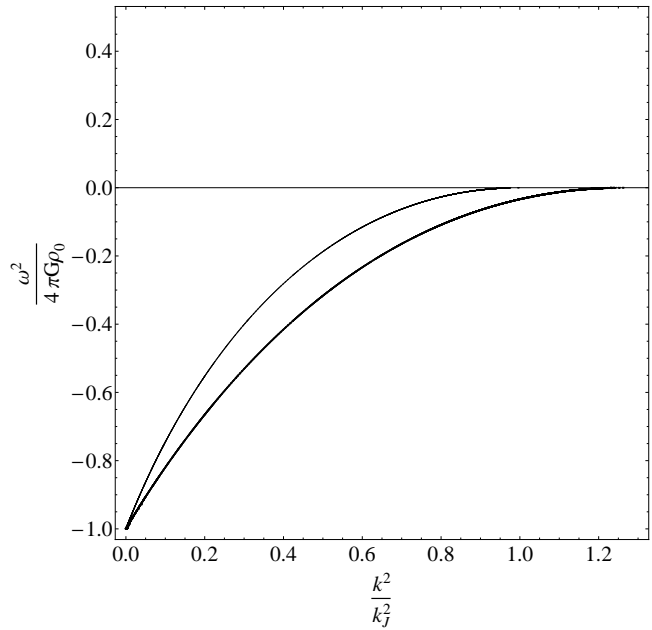}\\
  \caption{The bold line indicates the plot of the dispersion relation \eqref{eq:Results} in which we imposed
the value for $\alpha$ given by \eqref{alpha}. The thin line indicates the plot of the standard dispersion equation
\cite{BT}.}\label{Fig:Jeans}
\end{figure}

Our  task is now to check how the Jeans instability occurs in $f(R)$-gravity \cite{jeans}.
Let us  approach the Jeans instability with 
the  Poisson equations given by Eqs. \eqref{EQP} and \eqref{EQP2} after assuming the collisionless Boltzmann equation:
\begin{equation}
{\begin{array}{l}
\dfrac{{\partial f(\vec r,\vec v,t)}}{{\partial t}} + \left( {\vec v\cdot\vec \nabla _r } \right)f(\vec r,\vec v,t)  -\left( {\vec \nabla \Phi \cdot\vec \nabla _v } \right)f(\vec r,\vec v,t) = 0\,.
\end{array}}
\end{equation}
Then we have

\begin{equation}
 \nabla^2(\Phi+\Psi) -2 \alpha \nabla^4 (\Phi -\Psi)=16\pi G\int {{f}(\vec r,\vec v,t)} d\vec v{\mkern 1mu}
\end{equation}
\begin{equation}
 \nabla^2(\Phi-\Psi) +3 \alpha \nabla^4(\Phi-\Psi)=-8\pi G\int {{f}(\vec r,\vec v,t)} d\vec v{\mkern 1mu}\,.
\end{equation}
In the previous equations, we have replaced  $f''(0)$ with the greek letter $\alpha$.
As in standard case, we consider small perturbation to the equilibrium and linearize the equations. After
we write equations  in Fourier space so they became
\begin{eqnarray}\label{F1}
&&{ - i\omega {f_1} + \vec v\cdot\left( {i\vec k{f_1}} \right) - \left( {i\vec k{\Phi _1}} \right)\cdot\frac{{\partial {f_0}}}{{\partial \vec v}} = 0},
\\ \label{F2}
&&{  - {k^2}{(\Phi _1} +{\Psi _1}) - 2\alpha {k^4}({\Phi _1} -{\Psi _1}) = 16\pi G\int {{f_1}} d\vec v},
\\ \label{F3}
&& {  {k^2}({\Phi _1} -{\Psi _1}) - 3\alpha {k^4}({\Phi _1} -{\Psi _1}) =  8\pi G\int {{f_1}} d\vec v}.
\end{eqnarray}
Combining Eqs. \eqref{F2} and\eqref{F3}, we obtain a relation between $\Phi_1$ and $\Psi_1$,
\begin{equation*}
{{\Psi _1} =  \frac{{3 - 4\alpha {k^2}}}{{1 - 4\alpha {k^2}}}{\Phi _1}}
\end{equation*}
inserting this relation in Eq. \eqref{F2} and combining it with Eq. \eqref{F1}, we obtain the dispersion relation
\begin{equation}\label{newEQdispersion}
1 - 4\pi G\dfrac{{1 - 4\alpha {k^2}}}{{3\alpha {k^4} - {k^2}}}\int {\left( {\dfrac{{\vec k\cdot\dfrac{{\partial {f_0}}}{{\partial \vec v}}}}{{\vec v\cdot\vec k - \omega }}} \right)} d\vec v = 0.
\end{equation}
If we assume, as in standard case, that $f_0$ is given by \eqref{Maxwellian} and $\vec{k} = (k,0,0)$, one can write
\begin{equation}\label{NewEqDispersion2}
{1 + \frac{{2\sqrt {2\pi } G{\rho _0}}}{{{\sigma ^3}}}\dfrac{{1 - 4\alpha {k^2}}}{{3\alpha {k^4} - {k^2}}}\left[ {\int {\dfrac{{k{v_x}{e^{ - \frac{{v_x^2}}{{2{\sigma ^2}}}}}}}{{k{v_x} - \omega }}} d{v_x}} \right] = 0}.
\end{equation}
By eliminating the higher-order terms (imposing $\alpha=0$), we obtain again  the standard dispersion Eq. \eqref{eqDISP}.
In order to compute the integral in the dispersion relation \eqref{NewEqDispersion2}, we consider the same approach  used
in the classical case, and finally we obtain:
\begin{equation}\label{eq:Results}
{\begin{array}{l}
1 + \mathcal{G}{\dfrac{{1 - 4\alpha {k^2}}}{{3\alpha {k^4} - {k^2}}}} \left[ {1 - \sqrt\pi x e^{x^2}\left( {1 - {\rm{erf}}\left[ x\right]} \right)} \right] = 0,
\end{array}}
\end{equation}
where $x=\dfrac{{{\omega _i}}}{{\sqrt 2 k\sigma }}$ and $\mathcal{G}=\dfrac{{4G\pi {\rho _0}}}{{{\sigma ^2}}}$.
In order to  evaluate Eq. \eqref{eq:Results} comparing it with the classical one, given by Eq. \eqref{eqDISP}, it is very useful to
normalize the equation to the classical Jeans length showed in Eq. \eqref{length}, by fixing the parameter of  $f(R)$-gravity, that is
\begin{equation}\label{alpha}
    \alpha   =  - \frac{1}{{k_j^2}}=  - \frac{{{\sigma ^2}}}{{4\pi G{\rho _0}}}.
\end{equation}
This parameterization is correct because the dimension $\alpha$ (an inverse of squared length) allows us to parametrize  as in the standard case.
Finally we  write
\begin{equation}
\begin{array}{l}\label{instability}
\dfrac{{3{k^4}}}{{k_j^4}}+ \dfrac{{{k^2}}}{{k_j^2}} = \left( {\dfrac{{4{k^2}}}{{k_j^2}} + 1} \right) \left[ {1 - \sqrt\pi x e^{x^2}\left( {1 - {\rm{erf}}\left[ x\right]} \right)} \right] =0.
\end{array}
\end{equation}
The function is plotted in Fig.\ref{Fig:Jeans}, where Eq. \eqref{eq:Results} and the
standard dispersion  \cite{BT} are confronted in order to see the difference between $f(R)$ and Newtonian gravity.

As shown in the Figure \ref{Fig:Jeans}, the effects of a different theory of gravity changes the limit of instability. The limit is higher than
the classical case and the curve has a greater slope. This fact is important because  the mass limit value of
 interstellar clouds decreases changing the initial conditions to start the collapse. This feature could have dramatic effects for star and black hole formation.

\section{The Jeans mass limit in $f(R)$-gravity}
\label{dodici}

 A numerical estimation of the $f(R)$-instability length in terms of the standard Newtonian one can be achieved. By solving numerically  Eq. \eqref{instability} with the condition $\omega=0$, we obtain that the collapse occurs for
\begin{equation}\label{newlengthnumerical}
    k^2=1.2637 k^2_J.
\end{equation}
However we can estimate also analytically the limit for the instability. In order to evaluate the Jeans mass limit in $f(R)$-gravity, we set $\omega=0$ in  Eq. \eqref{NewEqDispersion2} and then
\begin{equation}\label{NewMass}
3{\sigma ^2}\alpha {k^4} - \left( {16\pi G{\rho _0}\alpha  + {\sigma ^2}} \right){k^2} + 4\pi G{\rho _0} = 0.
\end{equation}
It is worth stressing that the additional condition $\alpha<0$   discriminates  the class of viable $f(R)$ models: in such a case   we obtain stable cosmological solution
and positively defined massive states \cite{Capozziello10}.  In other words, this condition selects  the physically viable models allowing  to
solve Eq.\eqref{NewMass} for real values of k. In particular, the above numerical solution can be recast as 
\begin{equation}
    k^2=\frac{2}{3} \left(3+\sqrt{21}\right) \pi  \frac{G \rho }{\sigma ^2}.
\end{equation}
The relation to the Newtonian value of the Jeans instability is
\begin{equation}
    k^2=\frac{1}{6} \left(3+\sqrt{21}\right) k^2_J.
\end{equation}
Now, we can define the new Jeans mass as:
\begin{equation}\label{newMass2}
    \tilde{M}_J=6\sqrt{\frac{6}{\left(3+\sqrt{21}\right)^3}}  M_J,
\end{equation}
that is proportional to the standard Newtonian value. We will confront this specific solutions with some observed structures.

Before this comparison, some considerations are in order.
 Star formation is one of the best settled problems of modern astrophysics. However, some shortcomings emerge as soon as one faces  dynamics of  diffuse gas evolving  into stars and  star formation in  galactic environment.
One can deal with the star formation problem in two ways:
$i)$ we  can take into account the formation of  individual stars  and $ii)$ we can  discuss the formation of the  whole star  system   starting from interstellar clouds \cite{Mckee2}.
To answer these problems it is very important to study the interstellar medium (ISM) and its properties.The ISM physical conditions 
in the galaxies change in a very wide range, from hot X-ray emitting plasma to cold molecular gas, so it is very complicated to classify
the ISM by its properties. However, we can distinguish, in the first approximation, between \cite{carroll, kipp, dopita, Scheff}:
\begin{itemize}
\item {\bf Diffuse Hydrogen Clouds}. The most powerful tool to measure the properties of these clouds is the 21cm line emission of HI.
      They are cold clouds so the temperature is in the range $10\div 50$ K, and their  extension is up to  $50\div 100$ kpc from galactic center.
\item {\bf Diffuse Molecular Clouds}  are generally self-gravitating, magnetized, turbulent fluids systems, observed in sub-mm.  The most of the molecular gas is $H_2$, and the rest is CO. Here, the
      conditions are very similar to the HI clouds but in this case, the cloud can be more massive. They have, typically, masses in the range $3\div100 \,M_\odot$, temperature in $15\div 50$K and particle density in $(5\div50) \times 10^8$ m$^{-3}$.
\item {\bf Giant Molecular Clouds}  are very large complexes of particles (dust and gas), in which the range of the masses is typically $10^5\div 10^6\, M_\odot$ but they are very cold. The temperature is $\sim15$K, and the number of particles is $(1\div 3) \times 10^8$ m$^{-3}$ \cite{Mckee, Mckee2, Blitz, Blitz2}.
      However, there  exist also  small molecular clouds with masses $M < 10^4 M_\odot$\cite{Blitz2}.
They are the best sites for star formation, despite the mechanism of formation does not recover the star formation rate that  would be $250 M_\odot yr^{-1}$ \cite{Mckee}.
\item {\bf HII regions}. They are ISM regions with temperatures in the range  $10^3 \div 10^4$ K, emitting primarily in the radio and IR regions. At low frequencies,
       observations are associated to free-free electron transition (thermal Bremsstrahlung).
      Their densities range from over a million particles per cm$^3$ in the ultra-compact H II regions to only a few particles per cm$^3$
      in the largest and most extended regions. This implies total masses between $10^2$ and $10^5$ M$_\odot$ \cite{HII}.
\item {\bf Bok Globules}  are dark clouds of dense cosmic dust and gas in which star formation sometimes takes place. Bok globules are found within
      H II regions, and typically have a mass of about $2$ to $50$  M$_\odot$ contained within a region of about a light year.
\end{itemize}

\begin{figure}[!h]
  \includegraphics[scale=0.70]{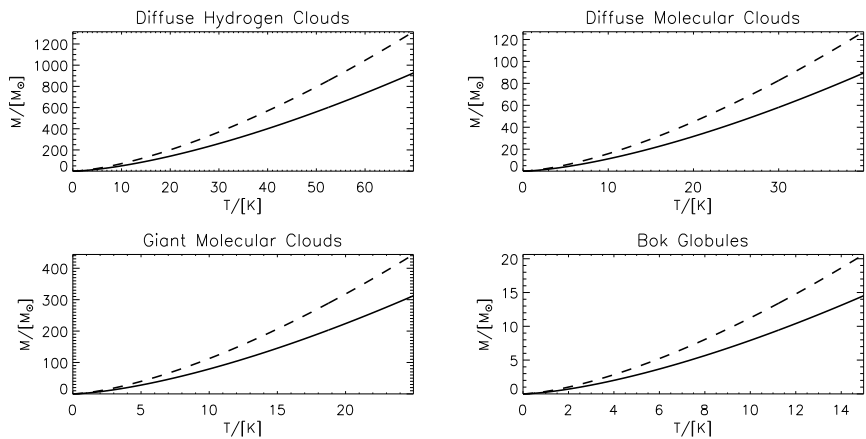}
  \caption{The  $M_J$-$T$ relation. Dashed-line indicates  the Newtonian Jeans mass behavior with respect to the temperature.
 Continue-line indicates the same for   $f(R)$-gravity Jeans mass.}\label{Fig:mass}
\end{figure}

Using  very general conditions \cite{carroll, kipp, dopita, Scheff, Mckee, Mckee2, Blitz, Blitz2,HII}, we want to show
the difference in the Jeans mass value between standard  and $f(R)$-gravity. Let us  take into account   Eq. \eqref{Mass} and  Eq.\eqref{newMass2}:
\begin{equation}\label{mass3}
{{{ M}_J} = \frac{\pi }{6}\sqrt {\frac{1}{{{\rho_0}}}{{\left( {\frac{{\pi {\sigma ^2}}}{{G}}} \right)}^3}} },
\end{equation}
in which $\rho_0$ is the ISM density and $\sigma$ is  the velocity dispersion of  particles due to the
temperature. These two quantities are defined as
$$\rho_0= m_{H} n_{H} \mu, \rm{\quad \quad} \sigma^2=\frac{k_BT}{m_H}$$
where $n_H$ is the number of particles measured in $m^{-3}$, $\mu$ is the mean molecular weight, $k_B$ is the Boltzmann constant and
$m_H$ is the proton mass. By using these relations, we are able to compute the Jeans mass for interstellar clouds and to show the behavior of the Jeans mass with the temperature. Results are shown in Tab.\ref{table1} and Fig.\ref{Fig:mass}
\begin{table}[!h]
\small{\begin{tabular}[c]{|l|ccccc|}
\hline
Subject			  &       T   &         n          &   $\mu$    &       $M_J$     &  $\tilde{M}_J$ \\
		          &      (K)  &  ($10^8$m$^{-3}$)  &            &    ($M_\odot$)  &  ($M_\odot$)   \\
\hline
Diffuse Hydrogen Clouds   &      50   &   	5.0   	   &     1  	&      795.13     &    559.68     \\
Diffuse Molecular Clouds  &      30   &   	50   	   &     2 	&      82.63 	  &     58.16	   \\
Giant Molecular Clouds    &      15   &   	1.0   	   &     2 	&      206.58 	  &    145.41    \\
Bok Globules  		  &      10   &   	100 	   &     2 	&      11.24      &      7.91   \\
\hline
\end{tabular}}
\caption{ Jeans masses derived  from Eq. \eqref{Mass} (Newtonian gravity)  and \eqref{newMass2}  ($f(R)$-gravity).} \label{table1}
\end{table}


\begin{center}
\begin{table}[!h]
\small{\begin{tabular}[c]{|l|cccc|}
\hline
     Subject	  &       T    &         n     &     $M_J$     &  $\tilde{M}_J$ \\
		          &      K  &  ($10^8$m$^{-3}$)  &    ($M_\odot$)  &  ($M_\odot$)   \\
\hline
GRSMC G053.59+00.04 & 5.97 & 1.48  &  18.25 &   12.85 \\
GRSMC G049.49-00.41 & 6.48 & 1.54  &  21.32 &   15.00 \\
GRSMC G018.89-00.51 & 6.61 & 1.58  &  22.65 &   15.94 \\
GRSMC G030.49-00.36 & 7.05 & 1.66  &  22.81 &   16.06 \\
GRSMC G035.14-00.76 & 7.11 & 1.89  &  28.88 &  20.33 \\
GRSMC G034.24+00.14 & 7.15 & 2.04  &  29.61 &  20.84 \\
GRSMC G019.94-00.81 & 7.17 & 2.43  &  29.80 &  20.98 \\
GRSMC G038.94-00.46 & 7.35 & 2.61  &  31.27 &  22.01 \\
GRSMC G053.14+00.04 & 7.78 & 2.67  &  32.06 &  22.56 \\
GRSMC G022.44+00.34 & 7.83 & 2.79  &  32.78 &  23.08 \\
GRSMC G049.39-00.26 & 7.90 & 2.81  &  35.64 &  25.09 \\
GRSMC G019.39-00.01 & 7.99 & 2.87  &  35.84 &  25.23 \\
GRSMC G034.74-00.66 & 8.27 & 3.04  &  36.94 &  26.00 \\
GRSMC G023.04-00.41 & 8.28 & 3.06  &  38.22 &  26.90 \\
GRSMC G018.69-00.06 & 8.30 & 3.62  &  40.34 &  28.40 \\
GRSMC G023.24-00.36 & 8.57 & 3.75  &  41.10 &  28.93 \\
GRSMC G019.89-00.56 & 8.64 & 3.87  &  41.82 &  29.44 \\
GRSMC G022.04+00.19 & 8.69 & 4.41  &  47.02 &  33.10 \\
GRSMC G018.89-00.66 & 8.79 & 4.46  &  47.73 &  33.60 \\
GRSMC G023.34-00.21 & 8.87 & 4.99  &  48.98 &  34.48 \\
GRSMC G034.99+00.34 & 8.90 & 5.74  &  50.44 &  35.50 \\
GRSMC G029.64-00.61 & 8.90 & 6.14  &  55.41 &  39.00 \\
GRSMC G018.94-00.26 & 9.16 & 6.16  &  55.64 &  39.16 \\
GRSMC G024.94-00.16 & 9.17 & 6.93  &  56.81 &  39.99 \\
GRSMC G025.19-00.26 & 9.72 & 7.11  &  58.21 &  40.97 \\
GRSMC G019.84-00.41 & 9.97 & 11.3  &  58.52 &  41.19 \\
\hline
\end{tabular}}
\caption{The name of  molecular clouds, the particle number density, the excitation temperature and the value of Jeans mass  are reported for  Newtonian and $f(R)$ case, respectively. 
This table is only a part of the catalog of molecular clouds reported in  \cite{duval}}\label{table2}
\end{table}
\end{center}

By using Eq.\eqref{newMass2} and by referring to the catalog of molecular clouds in Roman-Duval J. et al. \cite{duval}, we have
calculated the Jeans mass in the Newtonian and $f(R)$ cases. Tab.\ref{table2} shows the results. In all cases we note a substantial
difference between the classical and $f(R)$ value. We can conclude that, in $f(R)$-scenario, molecular clouds become sites where star formation is strongly
supported and more efficient.

\section{Discussion and conclusions}
\label{discuss}
Black hole solutions can be find out, in principle, in any Extended Theory of Gravity. Here, we have shown that exact solutions can be achieved in $f(R)$-gravity starting from symmetry considerations. In particular, the existence of Noether symmetries allow to find exact solutions. Such solutions can be eventually classified as black holes.

In particular,  we have shown that the Newman-Janis method, used  to derive rotating black hole solutions in GR, works also in $f(R)$-gravity.  In principle, it could be consistently applied any time a spherically symmetric solutions is derived. The method does not depend on the field equations but  directly works on the solutions that, a posteriori, has to be checked to fulfill the field equations.

The key point of the method is to find out a suitable  complex transformation which, from a physical viewpoint, corresponds to the fact that we are reducing the number of independent Killing vectors. From a mathematical viewpoint, it is  useful since allows to overcome the problem of a  direct search for  axially symmetric solutions that, in $f(R)$-gravity, could be extremely cumbersome due to the fourth-order  field equations. 

We have to stress that the utility of  generating techniques is not simply to obtain a new metric, but a metric of a new spacetime with specific properties as the transformation properties of the energy-momentum tensor and Killing vectors. In its original application, the Newman-Janis procedure transforms an Einstein-Maxwell solution (Reissner-Nordstrom) into another Einstein-Maxwell solution (Kerr-Newman). As a particular case (setting the charge to zero) it is possible to achieve  the transformation between  two vacuum solutions (Schwarzschild and Kerr).
Also in case of $f(R)$-gravity,  new features emerge by adopting such a technique.  In particular, it is worth studying how certain features  of spherically simmetric metrics, derived in $f(R)$-gravity, result transformed in the new axially symmetric solutions for black holes. For example,  considering the $f(R)$  spherically symmetric solution studied here, the Ricci scalar evolves as  $ r^{-2}$ and then the asymptotic flatness is recovered.
Let us consider now the axially symmetric metric achieved by the Newman-Janis method. The parameter $a\neq 0$ indicates that the spherical symmetry ($a=0$) is broken. Such a parameter can be immediately related to the presence of an axis of symmetry and then to the fact that a Killing vector, related to the angle  $\theta$, has been lost.
To conclude, we can say that once the vacuum case is discussed, more general spherical metrics can be transformed in new axially symmetric metrics  adopting more general techniques \cite{bk:dk}, that would allow us to have the best configurations to describe the  black hole dynamics.

A part the mathematical interest in finding out new solutions, such results could have remarkable applications, at least as toy models,  in the study of self-gravitating systems.
We have shown that the stellar theory as well as the  Jeans analysis of instability and collapse  can be dramatically altered if one adopt $f(R)$-gravity instead of GR.

The study has been performed starting from the Newtonian limit of $f(R)$-field equations. Since the field equations satisfy in any case the Bianchi identity,  we can use the  conservation law of energy-momentum tensor. In particular adopting a polytropic equation of state relating the mass density to the  pressure, we derive the \emph{modified Lan\'{e}-Emden equation} and its solutions for $n\,=\,0,\,1$ which can be compared to the analogous solutions coming from the Newtonian limit of GR. When we consider the limit $f(R)\,\rightarrow\,R$, we obtain the standard hydrostatic equilibrium theory coming from  GR. A peculiarity of $f(R)$-gravity is the non-viability of Gauss theorem and then the \emph{modified Lan\'{e}-Emden equation} is an integro-differential equation where the mass distribution plays a crucial role. Furthermore the correlation between two points in the star is given by a Yukawa-like term of the corresponding  Green function. 

These solutions have been matched with those coming from   GR and the corresponding density radial  profiles have been derived. In the case $n\,=\,0$, we find an exact  solution, while, for $n\,=\,1$, we used a perturbative analysis with respect to the solution coming from GR. It is possible to demonstrate that   density radial profiles  coming   from $f(R)$-gravity analytic models and close to those coming from GR are compatible. This result   rules out  some wrong  claims in the literature stating that $f(R)$-gravity is not compatible with self-gravitating systems. Obviously the choice of the free parameter of the theory has to be consistent with boundary conditions and then   the solutions are parameterized by a suitable "wave-length" $m\,=\,\sqrt{-\frac{1}{3f''(0)}}$ that should be experimentally  fixed.

The next step is to derive self-consistent numerical solutions of \emph{modified Lan\'{e}-Emden equation} and build up  realistic   star models  where further   values of the polytropic index $n$ and other physical parameters, e.g.  temperature, opacity, transport of energy,  are considered. Interesting cases are the non-relativistic limit ($n\,=\,3/2$) and relativistic limit ($n\,=\,3$) of completely degenerate gas. These models  are a challenging task since, up to now,  there is no  self-consistent,  final   explanation for  compact objects (e.g. neutron stars) with masses larger than Volkoff mass, while observational evidences  widely indicate these objects \cite{mag}. In fact it is plausible that the gravity manifests itself on different characteristic lengths and  also  other contributions in the gravitational potential should be considered for these exotic objects. As we have seen above,  the gravitational potential well results modified by  higher-order corrections in the curvature. In particular, it is possible to show that  if we put in the gravitational action  other curvature invariants also repulsive contributions can emerge \cite{stabile_2,cqg}. These situations have to be seriously taken into account in order to address several issues of relativistic astrophysics that seem to be out of the explanation range of the standard theory.

For this purpose,  we  have  analyzed the Jeans instability mechanism, adopted for star formation,  considering the
Newtonian approximation of  $f(R)$-gravity. The related Boltzmann-Vlasov system leads to  modified
Poisson equations depending on the $f(R)$-model. In particular, 
considering  Eqs.\eqref{EQP} and
\eqref{EQP2},  it is possible to get a new dispersion relation \eqref{eq:Results} where 
instability criterion results modified (see also \cite{poly}). The leading parameter is $\alpha$, i.e.  the second derivative of the specific $f(R)$-model. Standard Newtonian Jeans instability is immediately recovered for $\alpha=0$ corresponding to the Hilbert-Einstein Lagrangian of GR.
 In Fig. \ref{Fig:Jeans}, dispersion relations for Newtonian and a specific $f(R)$-model are numerically compared. 
 The modified characteristic length van be given in terms of the classical one. 

Both in the classical  and  in $f(R)$ analysis,
the system damps the perturbation. This damping is not associated to the collisions
because we neglect them in our treatment, but it is linked to the so called Landau
damping \cite{BT}.

A new condition for the gravitational instability is derived, showing  unstable modes with faster
growth rates. Finally we can observe the instability decrease 
in $f(R)$-gravity: such decrease is related to a larger
Jeans length and then to a lower Jeans mass. We have also compared the
behavior with the temperature of the Jeans mass for various types of interstellar
molecular clouds (Fig. \ref{Fig:mass}). In 
Tables \ref{table1} and \ref{table2} we show the results given by this
new limit of the Jeans mass for a sample of giant molecular clouds. In
our model the limit (in unit of mass) to start the collapse of an
interstellar cloud is lower than the classical one advantaging the
structure formation.  Real solutions for the Jean mass can be achieved only for
$\alpha<0$  and this result is in agreement with cosmology \cite{Capozziello10}.
In particular,  the condition $\alpha<0$ is essentials to have
a  well-formulated and well-posed Cauchy problem in $f(R)$-gravity  \cite{Capozziello10}.
Finally, it is worth noticing that  the Newtonian value is an upper limit for the Jean mass coinciding with $f(R)=R$.

This analysis is intended to indicate the possibility to deal with ISM collapsing clouds  under
different assumptions about gravity. It is important to stress that we  fully recover the 
standard  collapse mechanisms but we could also describe proto-stellar systems that escape the standard collapse model. On the other hand,  this is the first step to study star formation and physical black holes  in alternative theories of gravity (see also \cite{Cooney, Hu, poly, Chang}). From an observational point of view, 
reliable  constraints can be achieved from  a careful analysis of the proto-stellar phase taking into 
account magnetic fields, turbulence and collisions. Finally, addressing  stellar systems by this approach could be an extremely important to test observationally $f(R)$-gravity.

Moreover, the approach developed here  admits direct generalizations
for other modified gravities, like non-local gravity,
modified Gauss-Bonnet theory, string-inspired gravity, etc. In these cases,
the constrained Poisson equation may be even more complicated due to the
presence of extra scalar(s) in non-local or string-inspired gravity.
Developing further this approach gives, in general, the possibility to confront the
observable  dynamics of astrophysical objects (like stars) with predictions
of alternative gravities.

\section*{Acknowledgments}
The Authors wish to thank  I. De Martino, M. Formisano, S.D. Odintsov, A. Stabile for useful comments,  discussions and common works on the topics discussed here.

\end{document}